**A Pathway Selection Process for Dynamically Self-Organizing Systems**

**J. A. Sekhar** [1]

[1]College of Engineering, University of Cincinnati; Cincinnati, OH, USA
sekharja@ucmail.uc.edu

## Abstract

Self-organization creates new order and shifts sub-boundaries while reorganizing energy and entropy within a control volume. This article explores pathway selection and tests whether maximizing the entropy generation rate can predict process pathways. All entropy-generating processes distribute internal energy through temperature changes or structural responses, leading to new patterns or volume changes. Rapid self-organization, such as a supercooled liquid metal transforming into a solid, is a quasi-adiabatic process that tends to approach equilibrium or a steady state with respect to parameters like temperature. This is one of the main examples studied. Entropy generation relates to internal energy redistribution, either as work performed (called stored work) or as thermal energy stored or dissipated by the system. A system's resilience during and after self-organization is reflected in the emergence of measurable engineering properties, often from defect creation. In the analyzed examples, the entropy generation rate is maximized throughout the process, regardless of the work needed to create new boundaries. The examples also show a sigmoidal pattern in the transforming fraction or a similar intensive variable that tracks the amount transformed. Because the sigmoidal shape closely resembles well-known statistical distributions and their derivatives, these distributions and their derivative solutions are used to examine a broad class of processes within the context of entropy maximization. Self-organization is a dissipative process linked to pattern formation. The article discusses various patterns and shapes in physical systems, including grain size and morphology during thermo-mechanical deformation of crystalline solids, solid-liquid transformations, atmospheric effects, fluid-flow eddies, and patterned flight in birds that conserve and distribute energy under the framework of entropy-rate maximization. Morphological boundary limits are analyzed using the ratio of the energy dissipation rate to the entropy generation rate for several examples. This often results in similarities in morphological variations, as discussed for secondary dendrite arm spacing (curvature) during dendritic solidification and eddy size distribution in turbulent fluid flow. Regardless of the specific process, transformation processes can continue beyond an identifiable self-organizing phase, albeit with different time constants, thereby maintaining continuity and connectivity through entropy-generation maximization.

**Introduction:** Self-organization in systems results from internal communication, adaptation, and interactions among components. Complex systems that self-organize can spontaneously create new order, sub-boundaries, and repetitive features [1-165]. The processes often produce new morphologies and patterns to efficiently dissipate free energy. This article explores self-organized pathway decisions during transitions, focusing on entropy generation rate and relevant work events. The goal is to determine whether these paths are predictable and if they enhance resilience. During self-organization, the central system can alter its patterns in response to triggers, disrupting metastable states and inducing transitions. Self-organizing systems are resilient and develop new patterns or boundaries. In condensed matter, self-organized structures tend to be more adaptable and durable, with patterns associated with efficient resource utilization and energy storage within the system. The heating or cooling processes discussed in this article involve enthalpy and often gradients in chemical potential, temperature, defects, and pressure, thereby entailing entropy generation.

Studying rearrangements during transformation reveals the work performed on the system or the increase in its energy, as evidenced by boundaries, defects, and persistent gradients that persist beyond equilibrium. Figures 1 and 2 illustrate examples of ordering and observable patterns at various scales, highlighting sub-boundary regions across different levels of repetitive structure. Most spontaneous transformations are driven by the desire to decrease free energy within control volumes, but actual processes follow pathways that require additional energy to be distributed and, in some cases, a minimum energy input rate to overcome irreversibility. A system capable of spontaneous change is either metastable or unstable and needs a specific entropy production rate that only an evolutionary pathway can meet, as seen in nucleation, growth, and weather prediction. The entropy generation rate, emphasized by Onsager [18,40], is crucial for pathway selection, as discussed in the article. Typically, self-organized order is highly symmetrical, but the boundaries between sub-regions can be less so. While universal laws are symmetrical, real-world states may not be, leading to spontaneous symmetry breaking and the emergence of complex structures. Increased apparent order often results from energy release and redistribution (with boundary formation as is emphasized in this article), which raises overall entropy. For instance, a perfect crystal is highly ordered; melting causes atoms to move randomly, decreasing symmetry and increasing disorder, while maintaining some short-range order. Symmetry reflects the number of microstates associated with a given macrostate, complicating the understanding of equilibrium, which is often a local measure and does not capture the wide range of time scales over which different types of symmetry change. Spontaneous chemical reactions generally increase entropy and reduce symmetry, though some pathways may preserve it. The entropy generation rate per unit volume ($\dot{s}gen$) measures this [4,6,12-23, 52, 164,165], being zero at equilibrium and constant at steady state [4].

In most examples discussed here, overall volume changes are minimal but not necessarily zero, usually reflecting an amount equal to the inverse of the density difference between phases or regions with different symmetries during the initial and final states of a transformation in condensed matter. The main examples analyzed in detail relate to solidification processes. Key textbooks on solidification providing additional context on the physical and atomic processes involved in both equilibrium and non-equilibrium solidification, as well as the critical non-equilibrium defects mentioned here, are listed in [47-49]. The article cites published literature on solidification [47-49, 131, 132] and other transformations to validate experimental measurements and property trends, and to support the thermodynamic groupings discussed here for self-organized behavior [4, 5, 6, 19, 34, 44, 46 -146,150, 161-165].

The article is organized as follows: (1) An introduction to patterns and self-organization, emphasizing self-repeating morphologies, with subsections on common methods for categorizing and studying patterns or resilience; (2) Insights into irreversible processes, work, and the entropy generation rate in classical thermodynamics, including an introduction to stored work; (3) An analysis of validated instances of MEPR (maximum entropy production rate) for self-organization, entropy generation, and boundary defects, with examples from solidification, bird formation, fluid-flow, and solid-state transformations; (4) Simplified calculation methods using the MEPR condition for both steady-state and non-steady-state self-organization; (5) The significance of sigmoidal-shaped curves in sub-region formation; (6) Features of complex self-organization; and (7) Summary and Concluding remarks.

### 1.1 A Comparison of Analysis Methods for Self-Repeating Morphologies

Several complex objects demonstrate both resilience and the ability to generate (or self-generate), either dynamically or statically, across different scales of observation [4,52, 160, 161]. When self-organization occurs across multiple scales, the resulting patterns are known as fractal objects or self-similar patterns. They often appear rough or diffuse, similar to coastlines. A comparable roughness is also observed in other studies, for example, at solid-liquid interfaces or on plasma-treated surface features [52]. Objects are rarely linear, smooth, or free of flaws, a fact that becomes more noticeable at specific observation scales than at others [17, 65, 66, 67, 68]. Fractal objects enable the optimization of features at scales smaller than the observation scale [22, 44, 52]. In this article, we examine the conditions that lead to varying rates of sub-boundary creation. For all practical purposes, fractal dimensions are considered pathways that generate entropy. Fractal numbers associated with a pathway are thought to be linked to its entropy production rates. As a result, utility and resilience are interconnected properties across scales, with utility primarily involving integer powers and resilience sometimes involving non-integer powers. Self-repeating shapes, often called fractals, emerge or become visible through a process of self-organization. These patterns also appear in reaction-diffusion systems, where two or more components interact to create patterns [46-72,122]. Many of these patterns can be simulated by repeatedly applying simple rules, as in computer-generated fractal shapes such as the well-known Sierpinski triangles. In nature, these patterns arise from efficient, repetitive processes, including branching in trees, lightning, blood vessel formation, wave-like patterns in chaotic systems, star formation, and, at the molecular level, enzyme catalysis [20, 22, 25, 50, 51, 72, 77, 78]. These examples highlight similarities across different scales and time frames, as well as under oscillatory conditions, across various fields in which they are relevant. Self-similar systems and their associated organized behaviors are widely observed across disciplines [139]. This aspect is explored in more detail in Section 6. The exponent of a basic unit, like length, raised to a non-integer power, indicates a fractal measurement. However, self-similar patterns can be observed across different magnifications (length scales), often repeating within a control volume and in finer structures at smaller scales. Such structures may not have standard topological dimensions. Fractal dimensions and entropy are related [68,72], often connected morphologically by measurable features such as the area of a grain boundary [73], which we have described as a critical feature in certain types of self-organized systems. Fractal dimension provides a scale-independent measure of a fractal's complexity, similar to the entropy rate density [62].

Process pathways can be illustrated by the thermal routes shown in Figure 3. A liquid droplet can be cooled from a high temperature and solidified into a solid (a non-steady-state process) [76], or a crystal can be grown using the Czochralski (CZ) [4] method by pulling a liquid into a cooler region under steady conditions. Both techniques are essential in physical manufacturing from an engineering perspective. They involve complex, multicomponent, multiphase alloy transformations that result in a new, functionally useful, repetitive scale whose utility is tied to a specific length scale. For dynamic, local steady-state conditions, unlike in static macroscopic equilibrium, patterns need not be spherically symmetric or isotropic [5,47,48,49,52,53]. Under physical constraints, strong self-localized structures form via processes such as self-replication [50, 54, 55, 64]. Three types of spontaneous, self-organized control scenarios can be identified in self-organizing systems, illustrated in Figure 3. Figure 3(a) displays the time-temperature plot for a cooling liquid, which may or may not exhibit metastable states as it cools from a higher temperature to the external reservoir temperature. These include: (i) the system moving toward a thermodynamic equilibrium after a self-organizing event (path A, B, or C); (ii) reaching a new steady state or metastable equilibrium (path C) (as further discussed Section 3); and (iii) an oscillatory system that alternates between states while decaying very slowly (Figure 2). The first two are well exemplified by crystallization (crystal growth), either by cooling [47-49, 76] or by rapid pressurization [75,138], as shown in Figure 3(b). The third type is easily visualized in low- and high-temperature Belousov–Zhabotinsky reactions, which also influence the development of new high-temperature alloys [50, 51,52].

Diffuse interfaces are commonly employed to infer entropy generation and shape evolution (as discussed in more detail in Sections 2 and 3 below). As shapes are often characterized by Fractals, an assessment of their relationship is discussed below. Fractals and diffuse interfaces are closely connected because diffusion is a fundamental physical process that naturally creates fractal objects and interfaces [65, 150, 151, 152, 153, 154]. The connection arises from the inherent randomness and scale-invariance of diffusion phenomena, combined with the complexity of an interface. It has been shown [151] that fractal and entropy-producing diffuse interfaces (and sub-boundaries) are essentially different descriptions of the same structure within a complex system stretching across multiple volumes and part of a larger system. However, mathematically, fractals are not differentiable [65], which presents a challenge when applying fractal-based models of self-organizational transformation, unlike models that focus on the entropy-density generation rate. For physical systems, the units used to describe fractals require normalizing the entropy generation rate per unit volume. Nonetheless, it is widely accepted that complex systems can generate new entropy across different length and time scales, likely at varying rates depending on the driving conditions for self-organization [1-161]. For instance, clouds are estimated to have a fractal dimension of approximately 2.2; however, this does not provide a method for analyzing their dynamic behavior in rainfall prediction. Conversely, an approach that links the rate of entropy generation by treating the cloud as a diffuse interface enables modeling of climate events influenced by warming [44]. It is also important to recognize that fractal descriptions are not limited to static geometric patterns; they can also represent processes that occur over time [67]. Fractal-shaped structures tend to evolve to maximize their surface area for energy or nutrient exchange, as seen in the branching patterns of trees [65, 67]. When viewed as diffuse interfaces, the conditions required to describe multibranch dendritic shapes or highly intricate cellular patterns are those in which the highest entropy-generation rates per unit volume are associated with significant reorganization at smaller scales [52].

An increase in fractal dimension signifies more complex structures, such as additional dendritic branches or hierarchical formations. In terms of entropy generation rate, finer dendrites are associated with higher entropy generation rates [4]. However, suppose the demand for the entropy generation rate per unit continues to increase under imposed process conditions. In that case, morphology can evolve from finely spaced cells to a high-defect plane front, and eventually into a glassy state. This is shown in [52], where a plot of the entropy generation rate per unit volume versus the scale of dominant characteristic events illustrates how entropy generation shifts to smaller characteristic pattern-repeatability volumes. When the coefficient of friction is a key property for wear optimization, both references [52] and [69] demonstrate a non-linear relationship between the ratio of roughness and a measure of self-similar wave behavior. Non-dimensional properties, such as the coefficient of friction, relate to ratios like (Roughness/Auto Correlation Lengths), which are independent of scale [69].

This article applies the maximum entropy production rate (MEPR) postulate to analyze pathways leading to self-similar regions within self-organizing volumes. During liquid-to-solid transformations, solidified boundaries between similar regions are high-entropy areas, while different pathways can produce high-hardness lattice or sub-boundary regions, depending on the energy distribution. These pathways enhance resilience by affecting dislocation motion and crack propagation. Although not examined here in any detail, it is known that bacteria shape their forms through internal cytoskeletons and external cell walls, thereby increasing resilience through precise growth [160, 161]. Extensive published literature indicates that the maximum entropy generation rate serves as a pathway indicator of spontaneous self-organization at microscopic levels, influencing properties at larger scales [4, 6, 11, 12, 47, 49, 40, 41, 50, 51, 52, 53, 54, 55, 62, 70, 74, 161]. Figure 1(a-e) shows self-similar structures in frozen solids and cloud clusters at various magnifications, illustrating variations in boundaries and scales. Boundaries may include glassy phases within crystalline matrices. Dynamic, self-organized systems store energy and entropy in patterns called sub-boundary patterns. Repetitive oscillatory structures, like Belousov-Zhabotinsky patterns, form bands seen in animals (e.g., zebras), with macro- and micro-scale structures demonstrated in Figures 2(a-e). These patterns reflect complex self-organizing, oscillatory processes before reaching steady states. Striations in fatigue cracks are linked to microstructural differences that influence fatigue resistance, whereas zebra stripes and similar patterns aid pathway assessments and may serve functions such as individual signatures for identification and the optimization of other properties, which remain underexplored.

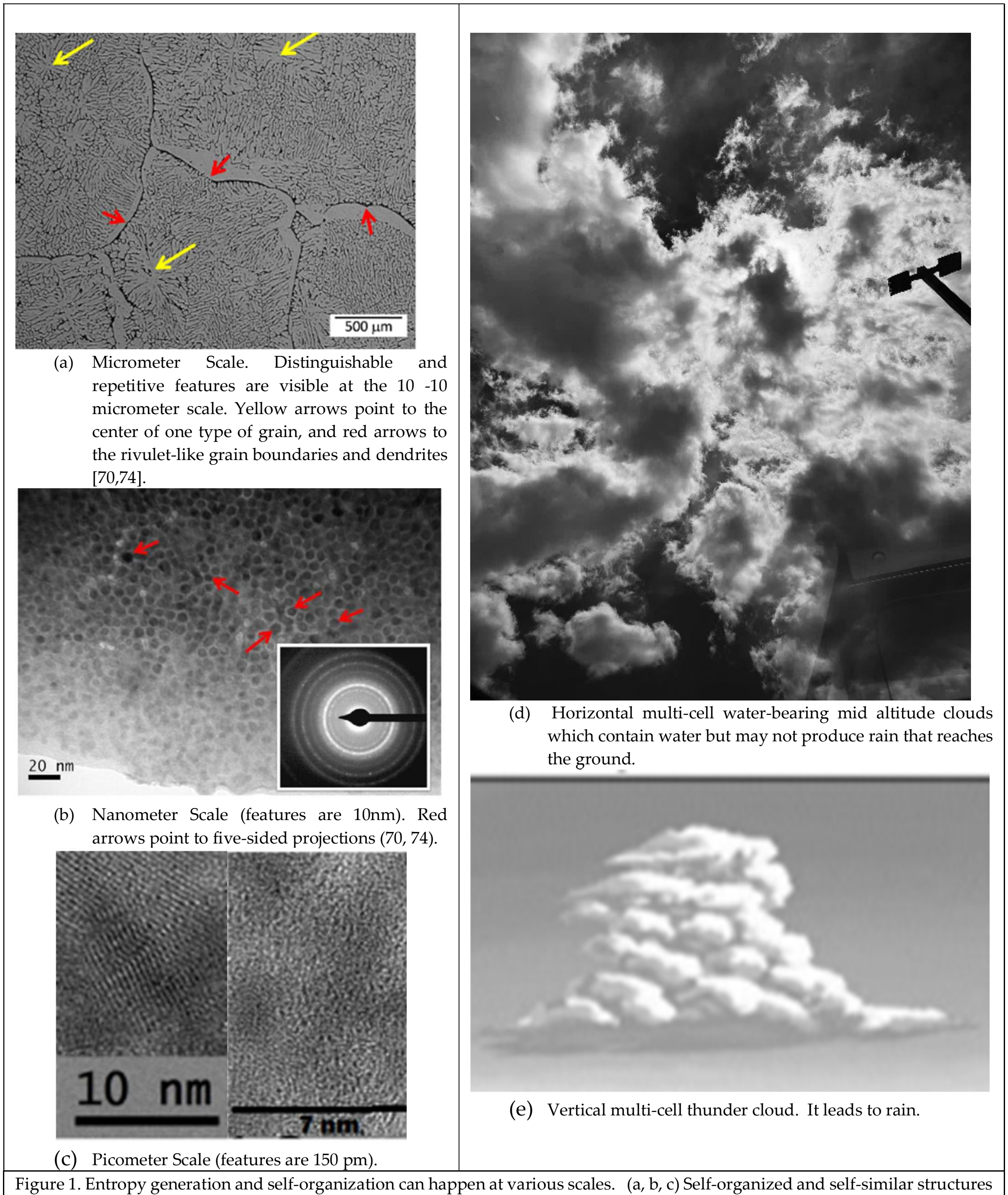

(a) Micrometer Scale. Distinguishable and repetitive features are visible at the 10 -10 micrometer scale. Yellow arrows point to the center of one type of grain, and red arrows to the rivulet-like grain boundaries and dendrites [70,74].

(b) Nanometer Scale (features are 10nm). Red arrows point to five-sided projections (70, 74).

(c) Picometer Scale (features are 150 pm).

(d) Horizontal multi-cell water-bearing mid altitude clouds which contain water but may not produce rain that reaches the ground.

(e) Vertical multi-cell thunder cloud. It leads to rain.

Figure 1. Entropy generation and self-organization can happen at various scales. (a, b, c) Self-organized and self-similar structures in cast (frozen) solids in a Cu-Sn alloy [70,74] at various scales of magnification. Figures (d, e) show entropy-generating non-equilibrium clouds with various scales of repetitive self-similar patterns [44]. The picture in 2(d) shows a streetlamp for scale assessment within a viewing angle of ~20 degrees. See reference [4] for the contextual assessment of self-similarity in the development of self-similar groups of rainclouds and their similarity to the grain and other microstructural features in a metallic crystalline object.

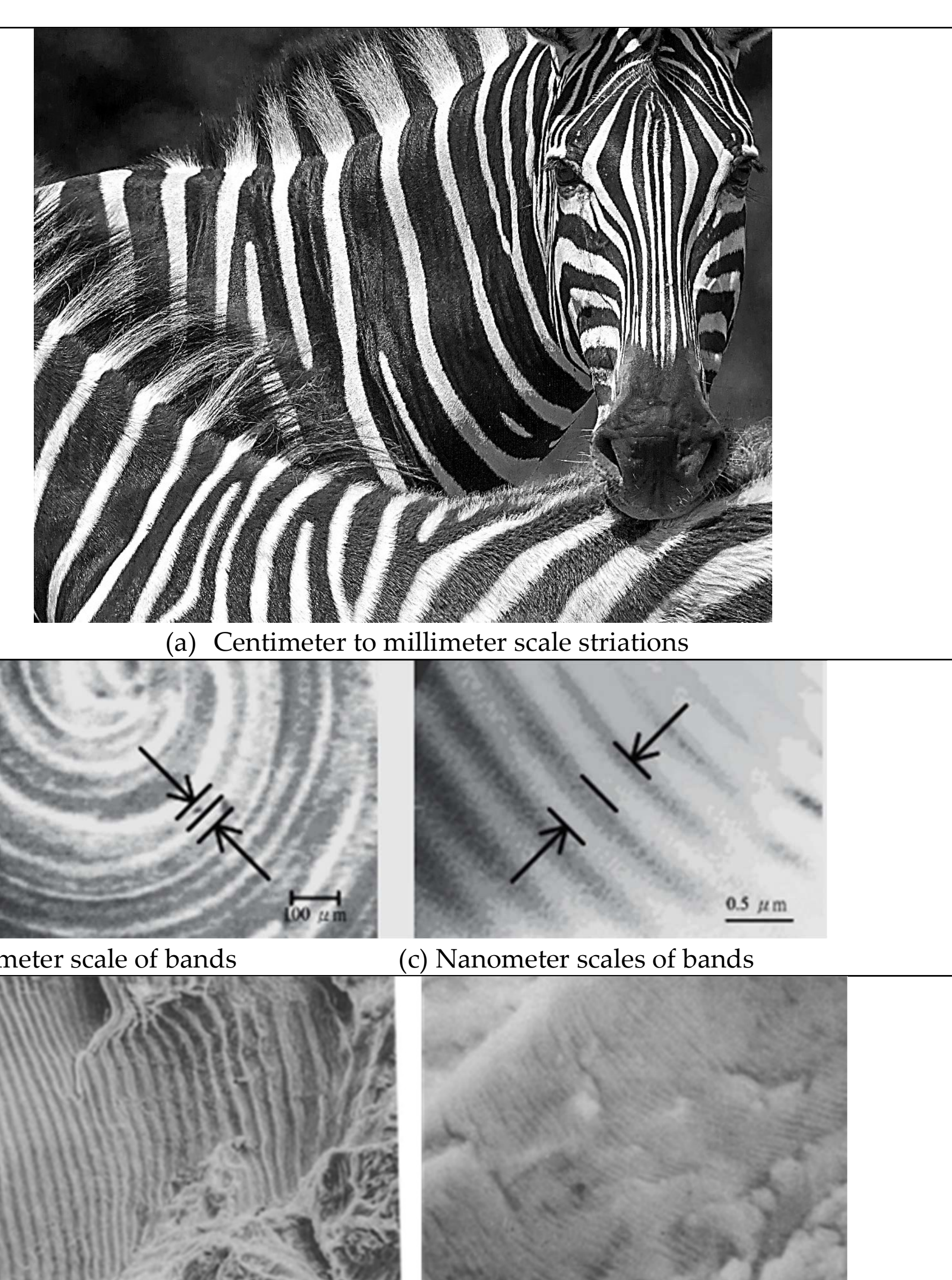

(a) Centimeter to millimeter scale striations

(b) Micrometer scale of bands (c) Nanometer scales of bands

(d) Micrometer scale of Fatigue Striations (e) Nanometer Scales of Striations

Figure 2, Millimeter, Micrometer, and Nanometer scale for Banded Self-Organizing (oscillating structures) and striations in Condensed Matter. Adapted from Reference [50, 51, 64, 79, 80].

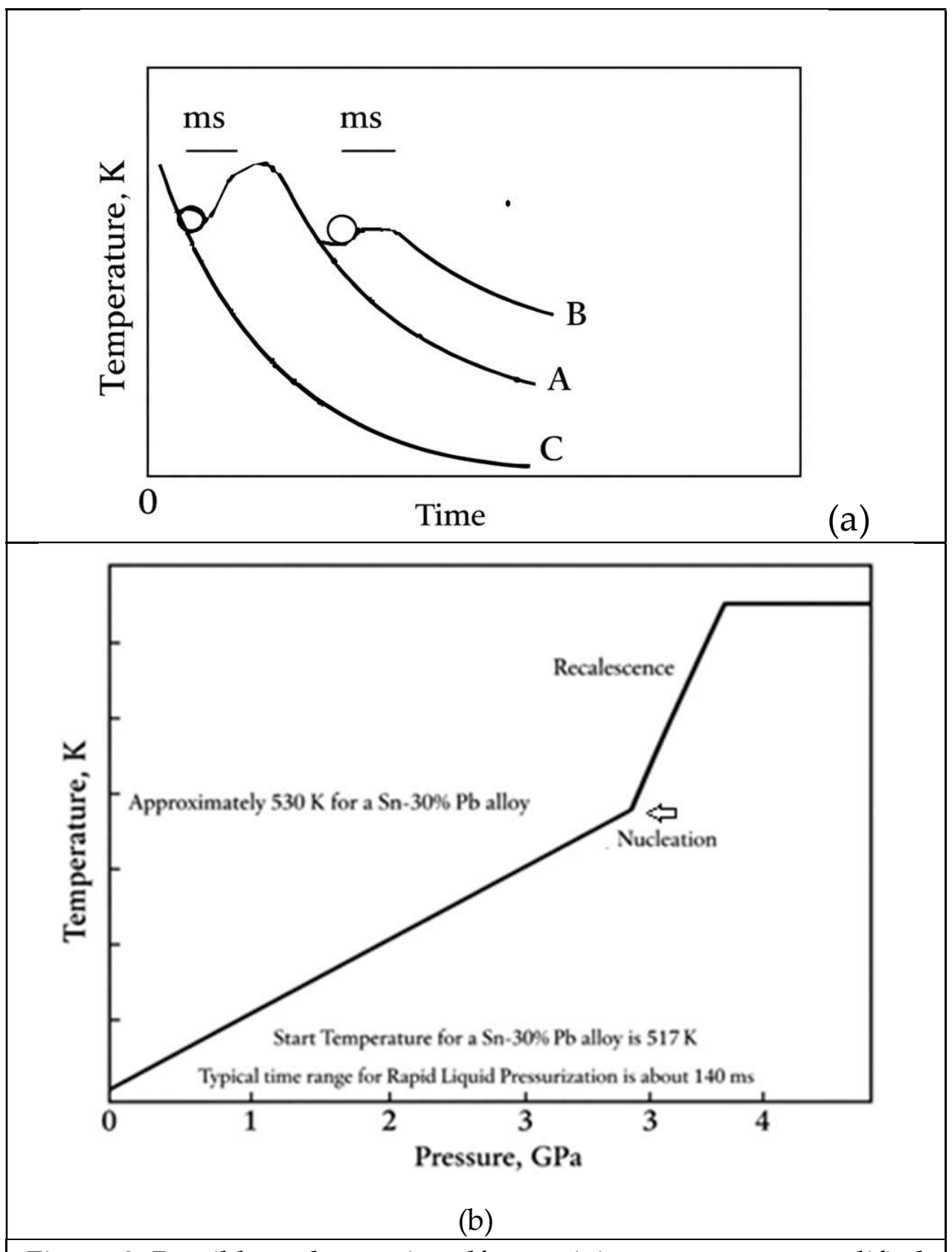

Figure 3. Possible pathways in self-organizing systems, exemplified by liquid droplet undercooling, which is discussed in detail in Section 4 of this article. Process pathways and potential metastable equilibria are shown in Figure 3(a) for Cooling and Recalescence [48, 49, 76], and in 3(b) for Rapid Liquid Pressurization [75, 138, 143]. Whenever recalescence occurs, it is rapid, lasting about 100 ms (milliseconds) [76] for metallic systems. Paths A and B [143] represent one or many metastable states, respectively. Path C shows no identifiable metastable states, and there is no intermediate heating from recalescence. After recalescence, following a metastable event, the decay toward equilibrium (path A) is usually exponential decay unless another metastable equilibrium is encountered (Path B). Resting in metastable states is illustrated by an analogy: with a ball at rest in a valley.

### 1.2 Utility and Resilience.

Self-organization leads to order and resilience, which have contextual meaning [141-161]. Resilience can be defined in various ways, as discussed below, and its significance can vary with the scale of analysis. In this article, we examine resilience at different scales that can result from self-organization and relate it to the energy processes underlying this phenomenon. For condensed materials, most experimentally measured engineering properties, such as yield strength or fracture toughness, and many others are, in a sense, measures of their resilience in a contextual sense discussed below (as opposed to fundamental thermodynamic properties like specific heat or thermal expansion), Material constants (engineering properties) either contain units of time (such as thermal conductivity) because they describe processes that inherently involve change, rate, or a delay over time, or they do not have time units if they describe static properties or instantaneous relationships between physical quantities.

Resilience is a property that enables disturbances to be dampened, rather than amplified. Turbulent flows are often described as "resilient" (or more commonly, "stable" or "robust" once established) because of their inherent ability to rapidly dissipate disturbances and maintain their overall chaotic, mixed state through internal mechanisms. It can be argued that properties such as the hardness of an alloy [43, 44, 47, 48, 49, 50, 69, 70, 74-87, 131, 132] reflect aspects of its usefulness, a term enabled by resilience. A material's stress intensity factor for resisting spontaneous crack growth [70, 79, 132] or wear [52, 69, 70, 74, 131, 132] could serve as a measure of its resilience. However, it is noteworthy that hardness also indicates resistance to dislocation movements at sub-boundaries or by precipitates, so both hardness and fracture resistance could be somewhat synonymous with different scales of measurement, but not necessarily so, except under certain conditions. Hardness of the lattice (the sub-regions) could result from Peierls stress for lattice dislocations, which represents the fundamental measure of the intrinsic lattice resistance to dislocation motion. Its value is unique to each material, depending on its crystal structure and bonding characteristics. Resistance to crack growth may be inversely related to lattice hardness unless sub-boundary regions become involved. Often, resistance to degradation or external stress must overcome the periodic potential of organized patterns, such as lattices or nanoprecipitates within the lattice or at sub-regions also called subboundaries [8, 85, 86]. Hardness is measured in Joules per cubic meter ($J/m^3$) and depends on grain or dendrite size [47, 48, 49]. Sometimes, hardness is also expressed in terms of lattice friction and a Hall-Petch constant with units of $J/m^2$. Fracture toughness, which indicates a material's resistance to crack growth, is measured in $J/m^2$ (80). These measures describe the resistance to dislocations or cracks moving smoothly across periodically varying potential barriers. However, fracture resistance is also expressed as energy per unit area, like hardness. Peierls stress, which resists dislocation movement, is measured in Pa ($N/m^2$) or sometimes as a non-dimensional number when divided by the shear modulus. A material with low Peierls stress, facilitating easier dislocation motion, is generally better at blunting cracks and thus more resilient or tougher. Conversely, high Peierls stress, as seen in ceramics, resists dislocation motion, making such materials more brittle and less tough. The specific wear rate has units of $m^3$/J (chalk: 10^-12 $m^2$/N to diamond: 10^-19 $m^2$/N) [69]. Graded or inhomogeneous materials and properties, such as viscoelasticity, involve non-integer powers of material properties. Fracture toughness varies with orientation in brittle materials, whereas hardness remains essentially isotropic when measured on large quasi-crystals with oriented sub-boundaries [124, 131, 132]. The phenomenon of material creep occurs when a material deforms under its own weight or under anisotropic loads, typically at high temperatures, near but below its melting point or the point of bond breakage. Here, increasing the number of boundaries increases the creep rate (strain rate), whereas fracture toughness decreases. Creep, unlike resistance to dislocations or cracks, could be an event that requires resilience during a self-organizing process (i.e., continuous feedback mechanisms are active). During creep, grain boundaries slide past each other, and in some cases, they may migrate toward planes of maximum shear stress, thereby changing the orientations of grains and grain boundaries. This reorientation is a key component of the deformation process and may also involve vacancies. This unusual phenomenon is discussed in Sections 4 and 5, as well as in references [18, 47, 48, 130, 131, 132].

We thus recognize that resilience is a term associated with the system's ability (or resistance) to withstand energy, i.e., the energy required per unit volume or area, and that a more resilient material withstands a more energetic disturbance (in the proper context of observation). Both heat- and work-based degradation can weaken the defenses of a periodic, ordered barrier in a similar manner, whether for an army facing an onslaught of intruders or for opposition on a more microscopic scale to dislocation movement that can cause yielding or growth in a crystalline lattice [70-86]. It should also be recognized that although the units of Hardness and Energy density are the same, they may contextually represent different types of physical phenomena and different types of contextual resistance.

This article examines the entropy generation rate as a key variable [4,5] in the formation of pathways and sub-boundaries during self-organization. The pathway selection influences the morphological scale of the self-organized pattern (4). In this way, the entropy generation rate is linked to measurable engineering properties because it affects pattern formation, which in turn affects these properties [4, 47-49, 79, 80, 86]. The article uses a classical thermodynamic approach, based on Onsager's work [18, 40], to address flux entropy generation arising from the irreversible transport of measurable quantities within a system. Irreversible transport refers to phenomena driven by energy, matter, pressure, and charge gradients. Another aspect of irreversible entropy generation is the configurational entropy, a component of a system's total entropy associated with the number of possible spatial arrangements (microstates) and their influence on the energy distribution. Entropy generation involves changes in these arrangements or the creation of new active microstates.

Several published studies highlight the importance of comparing a process to a reversible one in terms of the thermodynamic uncertainty of interactions (in a canonical sense) for self-organizing systems [61, 82, 83, 84]. Physical processes can be described by stochastic models, such as Markov chains and diffusion processes, and entropy production can be defined probabilistically for these models. As seen in many phase-field studies of liquid-to-solid ordering and morphological evolution, changes in an order parameter are used to infer the system's dynamics. The uncertainty per unit of work output can measure the efficiency of ordering during self-organization [61], relating it to a grouping like [(1/R)(δS(config)/δW)] where T, δS(config), R, and δW stand for temperature, configurational entropy (a form of entropy generation), the universal gas constant, and work, respectively. However, this grouping is somewhat awkward because the path should ideally be associated with a rate, which is difficult in probabilistic or stochastic models. Nonetheless, predictions [82,83,84] suggest maximum efficiency of ordering at critical temperatures, thus implying a maximal tendency for ordering based on the Gouy-Stodola theorem in reference [81], which states that the rate of exergy destruction (lost work potential) in a thermodynamic process is directly proportional to the rate of entropy generation during that process. Additionally, the maximum entropy generation rate or power requirements for various scales of subregion volumes involved in self-organizing transformations can be modeled by analyzing energy, matter, or configurational entropy fluxes, respectively [4, 18, 40, 52, 84].

## 2. Insights into Irreversible Processes, Work, and the Entropy Generation Rate in Classical Thermodynamic Formulations.

A system (control volume) can spontaneously transition from one equilibrium (or metastable) state to another. No real process path is reversible. A simple example is a hot object that cools spontaneously by releasing energy to the environment (the sink) via an irreversible heat-transfer process. Any irreversible process increases the entropy of the universe due to entropy generation. The presence of gradients in temperature, chemical potential, or pressure generates a rate of entropy ($\dot{S}_{gen}$), making the process path thermodynamically irreversible [18, 39, 40, 41, 40,41,42, 43, 71].

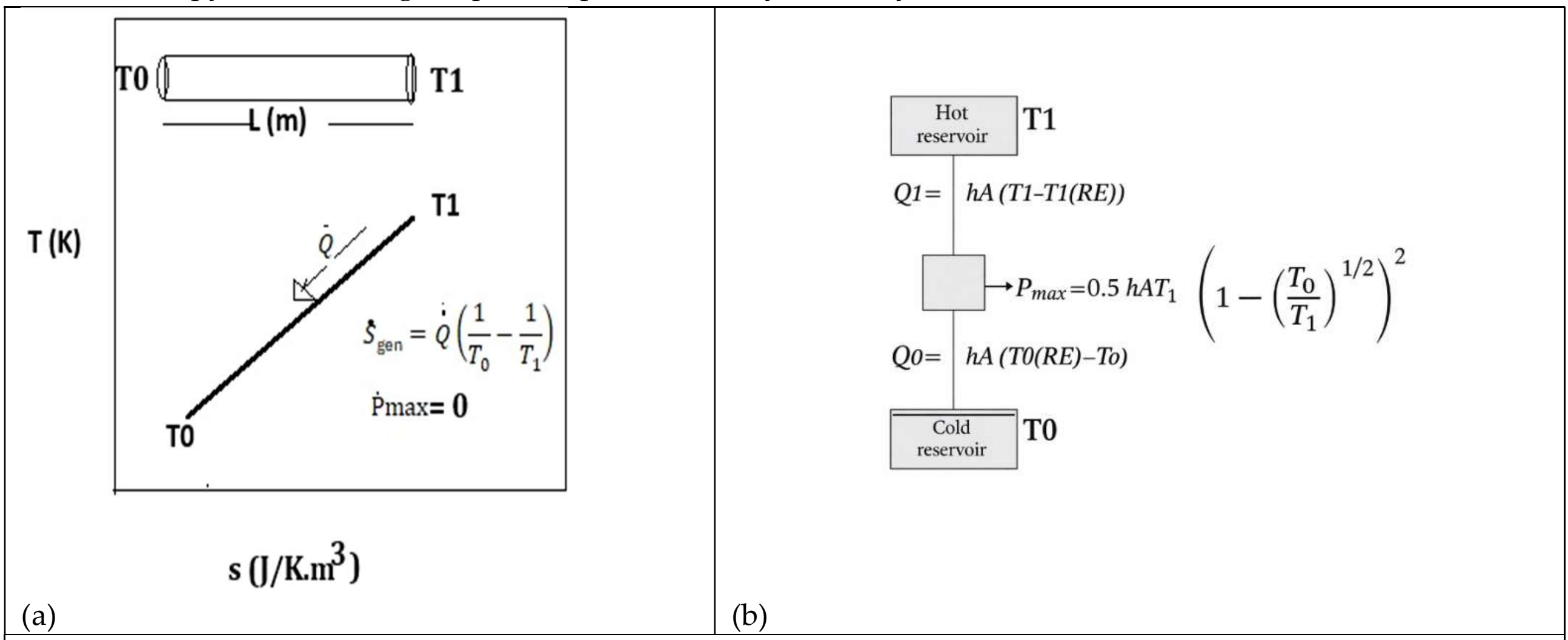

Figure 4 (a). A diagrammatic description of the temperature and entropy density, the heat flow rate between two reservoirs, entropy generation rate, and power rate in a bar with length $L$, cross-sectional area $A$, and thermal conductivity $\kappa$. The heat flow rate is $\dot{Q}$ (J/s) from T1 to T0 (with T1 > T0), at steady state, where both temperatures are reservoirs. This is a steady-state, irreversible pathway defined by local equilibrium conditions. The bar is adiabatic except at its two ends. Figure 4(b) shows a reversible heat engine operating between T1 and T0, yielding a maximum power output Pmax that depends on the heat-transfer coefficient (J/m²K) and the contact area A with the thermal reservoirs.

Assume that heat flows in a bar with length $L$, cross-sectional area $A$, and thermal conductivity $\kappa$, placed between two reservoirs $T_1$ and $T_0$ ($T_1$>$T_0$). A linear gradient is established (Figure 3), especially when the temperature difference is small. The heat flow rate, $\dot{Q}$ (J/s), at steady state is constant; the subscripts 1 and 0 denote the high-temperature and low-temperature reservoirs, respectively. Local equilibrium conditions prevail. The rate of entropy generation in the bar $\dot{S}_{\text{gen}}$ is [33, 71]:

$$\dot{S}_{\text{gen}} = \dot{Q}\left(\frac{1}{T_0} - \frac{1}{T_1}\right) \tag{2.1}$$

This new entropy also flows down the temperature gradient along with the thermal energy in the bar and is delivered to the reservoir at T0 once the steady state is established. Because of a linear temperature gradient, $\dot{Q}$ can also be written as:

$$\dot{Q} = \kappa \frac{A}{L}(T_1 - T_0) \tag{2.2}$$

Giving the entropy generation rate as,

$$\dot{S}_{\text{gen}} = \kappa \frac{A}{L} \frac{(T_1 - T_0)^2}{T_1 T_0} \tag{2.3}$$

The Carnot efficiency limits the maximum work that can be done between two temperatures. The maximum possible power, Pmax, available from a heat engine operating between $T_1$ and $T_0$ for a heat transfer rate $\dot{Q}1$ is

$$\text{Pmax(ideal)} = \dot{Q}1\ (T_1 - T_0)/T_1 \tag{2.4}$$

Thus, although the physical situation in Figure 4(a) is at a steady-state condition, it is an irreversible pathway defined by the local equilibrium conditions (The local equilibrium condition is required for the heat flux to be defined and for an experimentally measurable (verifiable) linear temperature gradient to be established).

$$\dot{S}_{\text{gen}} = \frac{\text{Pmax(ideal)}}{T_0} \tag{2.5}$$

Regardless, even for an idealistic heat engine operating between reservoirs $T_1$ and $T_0$, the maximum power output, Pmax (J/s), must include a heat transfer coefficient between the corresponding temperatures of the heat engine, $T_1$(RE), and $T_0$(RE), where RE stands for Reversible Engine. Figure 3(b) illustrates a reversible heat engine between two thermal reservoirs, yielding a maximum power (Pmax) that depends on the heat transfer coefficient (J/m² · K ·s) and the contact area (A) with the thermal reservoirs [71].

$$\text{Pmax} = 0.5\ hAT_1 \left(1 - \left(\frac{T_0}{T_1}\right)^{1/2}\right)^2 \tag{2.6}$$

*In the ideal case (i.e., when no entropy is generated)*

$$\dot{Q}0 = \dot{Q}1 - T_0 Pmax\ (ideal) \tag{2.7}$$

*In the realistic case*

$$\dot{Q}_0 = \dot{Q}_1 - 0.5\ hAT_1 \left(1 - \left(\frac{T_0}{T_1}\right)^{1/2}\right)^2 \tag{2.8}$$

Because $T_0 < T_1$, Equation (2.7) indicates that a realistic Pmax is always less than the ideal Pmax(ideal), shown in Equation (2.4), for a fixed energy (heat) rate extracted from the hot reservoir. The corresponding entropy generation, which effectively reduces the available power from the ideal Carnot amount, is transferred with the heat to the cold reservoir—that is, an energy rate sent to the cold reservoir exceeds what would have been transferred if no entropy had been generated. This also reflects work performed on the system to maintain the reversible-cycle processes shown in Figure 4(b).

Equation (2.5) indicates the maximum entropy generation rate is operative because it is the highest work rate divided by the lowest temperature within the range. *Therefore, the rate of entropy production is always maximized in a steady-state simple thermal heat transfer process or heat engine.* Steady state is generally defined as a condition where local equilibrium exists, regardless of thermodynamic fluxes, temperature, pressure, charge, or chemical potential gradients. In other words, it occurs when macroscopic temperature and other thermodynamic properties can be defined and measured. Similar arguments about the rate of entropy generation were first presented in references (4, 5).

Note that in the example shown in Figure 4, the rod length remains constant. The rod must expand due to a positive coefficient of thermal expansion, with T0 as the reservoir temperature. Therefore, constraining it requires work on the system. Energy dissipation is the process through which work and energy convert into forms that can be transferred as heat during a process. It primarily represents energy loss through irreversible processes, such as friction, resistance, and other dissipative forces, thereby reducing the system's mechanical or functional energy over time. However, this dissipation can instead be thought of as stored work (along with heat release), which creates sub-boundaries and is therefore considered stored work (Sections 3 and 4).

Work can be done on the system due to non-ideality. This permits defining Tav in terms of $T_1$ and $T_0$, as was done for solidification [3], where the product (Tav..*Ṡgen*) is also possibly maximized, as shown in Equation (2.9) below, because the heat transfer coefficient (h), the energy transfer rate, Q̇1, and the temperatures $T_1$ and $T_0$ are fixed in this problem. The loss in work-potential is thus approximated by:

$$\int T.\dot{S}gen = Tav.\int \dot{S}gen = [\dot{Q}1(1\text{-}T0/T1) - 0.5\,hAT_1\left(1-\left(\frac{T_0}{T_1}\right)^{1/2}\right)^2] \quad (2.9)$$

Which, in a more general form, can be written as,

$$[Tav.\int \dot{S}gen\,]/[\dot{Q}1(1\text{-}T0/T1) = [1-\{w.\,h.\,A.\,T_1\left(1-\left(\frac{T_0}{T_1}\right)^{1/2}\right)^2/\dot{Q}1(1\text{-}T0/T1)\}] \quad (2.10)$$

The fraction of the maximum useful power that can be realized at ideal efficiency is 1. This can be plotted for various power-conversion processes, as shown in Figure 5, using Equation 2.10 (the value above in Equation 2.9 is for w=0.5). The fraction of useful power possible is a function of the Carnot efficiency and can be written as $f(z)/z=[(1-w)+2w1-z-w(1-z)]/z$, where, $z=(1-\frac{T_0}{T_1})$. The fraction is shown in Figure 5. The loss in efficiency is determined by the irreversibility variable, the heat-transfer coefficient (h), which has units of W m-2 K-1, like the entropy generation rate per unit area. This fraction is shown in Figure 5.

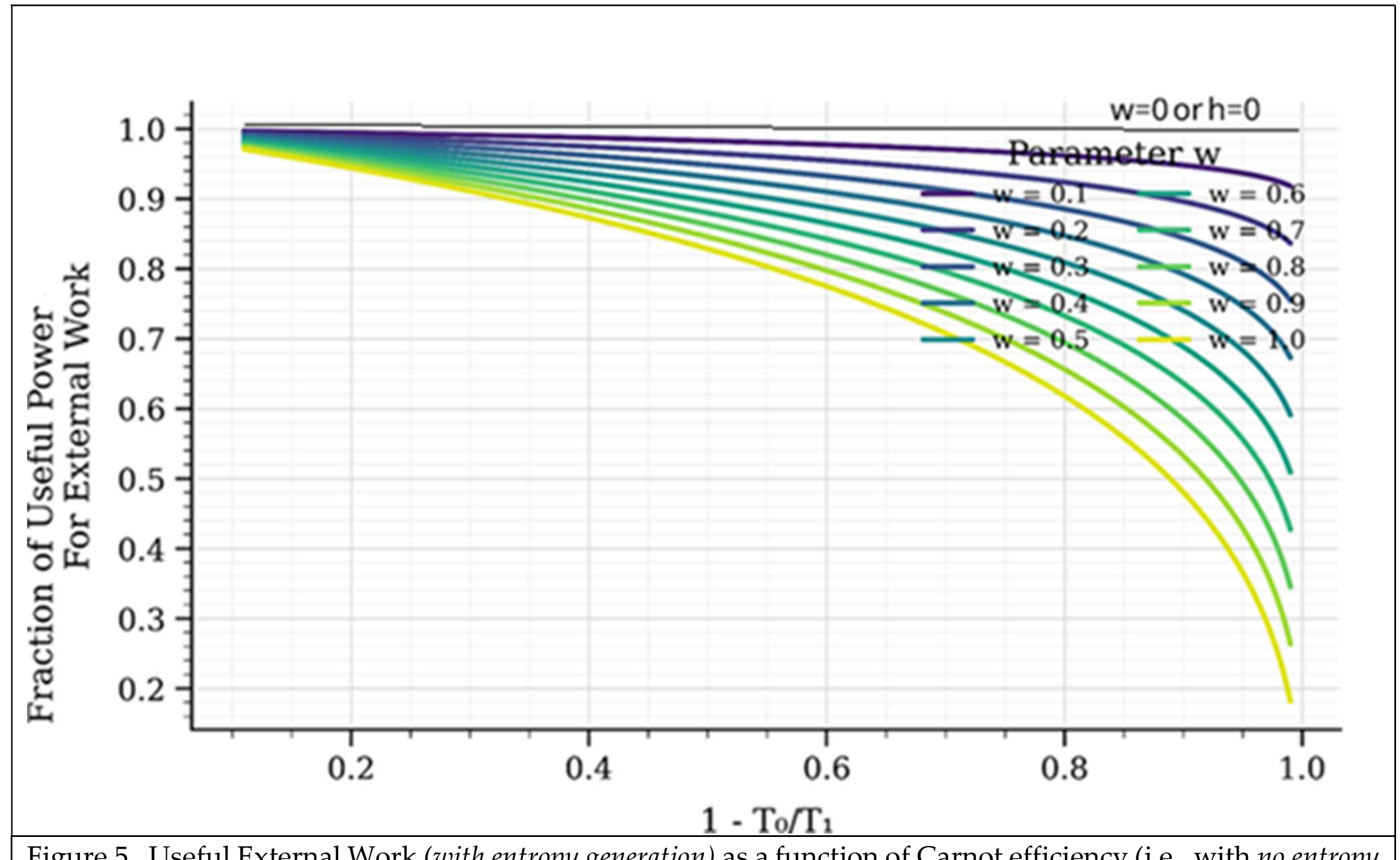

Figure 5. Useful External Work (*with entropy generation)* as a function of Carnot efficiency (i.e., with *no entropy generation*) under the assumption leading to Equation 2.10. $T_0$ is the lower reservoir temperature, and $T_1$ is the higher reservoir temperature. The scaling parameter w, as shown in Equation 2.10, effectively represents a finite heat-transfer coefficient (w.h) where w ranges from 0.1 to 1.

For Figure 5, we have assumed that Q̇1 = h. A.$T_1$. When h = 0, the fraction of useful work is 1 (Ideal Carnot efficiency) as per equation 2.10. The maximum *Loss in Power Efficiency*, as far as external work is concerned, occurs at the x-axis at 1, assuming Q̇1 = h.A.$T_1$. Following Figure 5, this condition for maximum loss occurs at a Tav established for $T_0/T_1 \approx 0$, which has a Pmax(ideal) of 1 (see the two-reservoir example discussed in Figure 4(b)). This result could indicate that if entropy generation rate is maximized (e.g., through spontaneous self-reorganization, as discussed in later sections of this article), the maximum work-conversion efficiency could be very low because of the irreversibility imposed by the heat transfer coefficient, h. Despite the loss of efficiency, internal work remains feasible. *Because the rate of entropy* *production is maximized in a steady-state simple thermal heat-transfer process or heat engine (Equation 2.5), this maximization could effectively reduce the power expended on useful (external) work. This is typical when a first-order transformation enables boundary formation, as will be seen in the later chapter but is also possible for mixed order reactions.*

The examples discussed above are cases in which the heat flow was spontaneous, the work done on the system was external, the rod was artificially compressed (an assumption that allows the use of thermal conductivity), or a heat engine operated between two temperatures, powered by an external energy source. Another way to think about this is to consider two equal masses initially at T1 and T0 that are brought to equilibrium by physical contact. If they were brought to equilibrium by touching each other (a condition that maximizes entropy generation), then the final temperature would be (T1 + T0)/2. If, instead, they were brought to an equilibrium temperature through heat transfer facilitated by a device that does work to prevent entropy generation, the equilibrium temperature would be different $\text{Tequilibrium} = \sqrt{T0.T1}$. The result fundamentally means work has been involved (mostly extracted as much as the Carnot principle will allow). However, in the second case, work could be done on the system to mitigate efficiency loss. Energy dissipation is traditionally defined as the process by which useful energy is transformed into a less useful, often unusable form, such as heat. This can increase the entropy at a specific location during a process. It's commonly regarded as the type of energy loss that results from irreversible processes, such as friction, electrical resistance, or other dissipative forces, which decrease the system's available useful energy (typically mechanical or electrical) over time. There is an analogy to electrical systems with a resistance and a reactance load, where the reactance of an inductor stores energy that is later released at a different phase of the AC cycle, thereby reducing the efficiency of heating with the resistor.

Note, however, that in the examples discussed above, a 'no entropy production (generation) pathway' results in an equilibrium temperature lower than that of the entropy-producing pathway. This is because the entropy generation pathway leads to an equilibrium or steady-state condition in which the internal energy may be higher or distributed differently, and the entropy is higher. This condition is often stated as follows: in most cases, adding energy to an object increases its entropy [162]. The distribution of this internal energy depends on the microstate, which may involve the formation of sub-boundaries or the emergence of new phases. However, energy commonly thought to be only dissipated as heat (from entropy generation) can, in principle, be partially converted into work within the system. It's important to note that work done on the system during a process can also be of the P.dV type of work (where P is pressure and V is volume), and it may be reversible. This is in addition to stored work, which converts heat into work (as shown in the fluid-flow examples in Section 3.3) and thus maximizes entropy generation. An irreversible process can produce heat or stored work. Understanding the work performed on the system is essential for analyzing patterns in entropy-producing systems, particularly for maximizing the entropy generation rate. This stored work and energy includes the energy transferred into the system and is either stored within its high-entropy sub-boundaries or within a lower-entropy lattice (subregions). Still, it can, in principle, be extracted during or after a spontaneous process with a start and end time. Some of this energy can be converted into work output with a delay relative to its creation, but doing so reduces the total work that can be extracted during the process (see Figure 5). It may also enhance energy efficiency, especially in thermal conservation processes, as illustrated by bird formation flight patterns discussed in the next section. The following sections examine whether the rate of entropy generation reaches a maximum and the implications of stored energy within sub-regions when no external work is performed or extracted, yet work is seemingly done on the system (the terminology "stored work" is used to distinguish it from external work).

## 3. Validations MEPR (Maximum Entropy Production Rate) for Self-Organization, Entropy Generation Rate.

A connection between a macroscopic principle and the microscopic features of self-organizing systems is established through the MEPR (Maximum Entropy Production Rate) principle, first introduced in the Gouy-Stodola theorem (see Section 1 above). This section offers a brief analysis of the MEPR and the known experimental validations. As MEPR pertains to morphological limits, a discussion of morphological features of turbulence and grain-shape features in sintering/recrystallization is included in this section.

As discussed in the introduction, pathways that maximize the entropy generation rate could provide a basis for explaining how complex, ordered patterns can form within a control volume during self-organization. While minimizing entropy production to zero is necessary to reach classical thermodynamic equilibrium in a closed system, open systems, at steady state conditions, must adjust their rate of entropy generation (4). The relationship between shape formation and entropy generation rates has also been studied by Martyushev [6] and Hill [38], among others, with examples from various fields [146-165]. Increasing the entropy production rate to its maximum within an open control volume effectively characterizes the emergence of patterns, shapes, and defects during steady-state transformation [4]. Patterns often optimize contextual objectives; for example, tree branching allows efficient water and nutrient distribution, tessellations optimize physical space, and lattice and crystal structures maximize strength-to-performance ratios. Coral growth and snowflakes may develop dendritic structures to maximize surface area. There are features of transformations or patterns that appear common across many MEPR validations to date. In entropy-generation-rate models, a diffuse-interface approach has proven helpful for accurately capturing morphological transitions [4,23,44,151]. With MEPR, explicit modeling of the role of boundary defects in driving entropy generation can be addressed. The key features of a diffuse interface include a finite thickness, a smooth transition in the order parameter or transformed fraction, and a phase-mixing process that is not readily visualized, even in clouds [44]. Within diffuse interfaces, the phases may not be strictly separated but are partially mixed, which more accurately reflects physical phenomena at the microscopic scale. The reaction occurring at the diffuse interface is commonly of mixed order [4]. According to the MEPR principle, a system's internal structure will evolve toward a state that maximizes its entropy generation rate ($\dot{S}_{gen}$) [4,6]. The straightforward, simple proof of MEPR presented in Section 2 above may not demonstrate that the MEPR hypothesis always holds. Regardless, the MEPR formulation has been successfully tested in more complex situations. Some of the experimental measurements and MEPR-based results are summarized in Tables 1 and 2 and discussed in detail in the subsections below. MEPR, a maximization principle, avoids the fine-tuning required in kinetic models with arbitrary-order parameters, as energy flow naturally aligns with efficiency goals. This is also demonstrated for heat engines (Section 2). Because the entropy generation can be calculated for a control volume using classical thermodynamic variables, it provides an additional maximization boundary condition to heat and work transfer problems. MEPR also enables seamless connections across different pattern scales.

### 3.1 Solidification and Diffusion Predictions

Bensah et al. [23,62] have shown that the entropy maximization rate is possibly a fundamental principle that describes transitions between microstructural morphologies, such as self-organized dendrites, cells, or grains that develop during solidification, and that it effectively predicts physical constants that have experimental validations [23, 52, 62]. Some of these measurements and calculations are summarized in Table 1. This involves mathematically accounting for the work and energy required to form new interfaces, such as those that form during solidification or wear [52]. MEPR incorporates terms that quantify the contributions of defects to the total entropy generation. These include the work rate (dW/dt) and defect entropy, which were introduced in the previous section as antiwork, a key process of self-organization. In reference 4, this is described as ($\omega_D/Tav$), where $\omega_D$ is the defect energy and Tav, as described in Section 2, is the average temperature at which the loss in efficiency is maximized. This term, along with an entropy term associated with chemical species segregation and two-phase mixing at the boundaries, contributes to the overall entropy generated. Similarly, it has been able to explain the lower friction coefficient resulting from self-organization during pairwise friction at the Hertzian regions between metallic contacts [52]. The importance of sub-boundary defects in driving entropy generation has been key in validating models of entropy generation rate. This involves mathematically accounting for the work and energy required to form new interfaces (sub-boundary regions), such as during solidification or wear.

### 3.2. Cloud Formation and Weather

Cloud formations and energy distributions across various weather events, such as wind, rain, and snow, can be assessed using MEPR boundary conditions [44]. Climate change is associated with an increase in energy flux through the Kolmogorov cascade discussed in Section 3.3. MEPR is useful in discussing the variations possible in the clouds that form in the troposphere [44]. The cloud cluster photograph (Figure 1d) was taken with an iPhone® 12 through a tinted window on a cloudy Saturday afternoon, August 23, 2025, at 2:10 pm. The sun is directly overhead, and a diffuse interface [44] exists between the dark and light areas within the clear sky. The camera captures blue and yellow hues, as well as dark, high-water-content regions. Despite a temperature of 28°C and a relative humidity of 54%, no rain occurred. This is inferred from the development of many clouds that formed horizontally rather than vertically (Figure 1(e)). Horizontal development (spreading rather than vertical growth) indicates a low-entropy-rate change [44]. It offers a useful conceptual and methodological tool for understanding the macroscale thermodynamic forces that drive thunderstorm activity. It can also aid in developing more reliable parameterizations of convective processes in large-scale climate models. MEPR has been used to analyze the behavior of both equilibrium and non-equilibrium clouds. Interactions between high-velocity winds and significant water-containing clouds can lead to multicellular thunderstorm formation. Lightning can also transfer entropy out of a cloud, thereby increasing the rate of entropy generation per unit time. For non-equilibrium thunderstorm cumulonimbus clouds experiencing the current tropospheric expansion rate, the upward velocity of thunderclouds, V(m/s), is related in the MEPR model [44] to the critical cloud thickness, $\xi c$ (m), and a constant Z ~0.05 (J kg-1s-1).

$$V_c^3 \sim Z.\xi_c \quad (3.1$$

### 3.3 Patterns Optimize Energy Savings and Infer Stability

Dynamic pattern stability at steady state or under equilibrium is often determined by symmetries and pattern formation [163]. This is well known in classical Vibro-shear experiments, where rotation and vibration together stabilize systems. For example, ball bearings self-organize in an orderly and stable manner in a rotational system, subject to vibration, at high angular speeds [141,163]. Similarly, the stability of shapes under steady-state conditions (i.e., constant energy exchange rate) is inferred from entropy-generation-rate calculations, which are analogous to equilibrium conditions that minimize energy and maximize entropy [4,38]. Stability is a property closely related to resilience, as discussed in Section 1.2. The symmetrical resting positions of ball bearings are governed by the conservation of angular momentum and the force balance that promotes well-defined organization. A similar organization occurs in birds during long journeys [5]. Stable patterns are also observed in vibrating plates [111] or other spherical objects [141,153], even biological analogs [164,165]. Organized behavior associated with standing waves [142] is often visible, although energy loss occurs over time, like decaying pattern-forming B-Z reactions [50,51,52].

During long migratory flights [5], a V-shaped formation is often observed at high altitudes, as shown in Figure 6. Modeling the complexities of wing mechanics, heat production, fluid dynamics, and heat dissipation is challenging. However, a straightforward energy and entropy balance model can be applied, assuming that the velocities are approximately equal on average across the control volume boundary. Flight angles, which align with experimental data using MEPR boundary conditions [5], are presented in Tables 1 and 2. Energy consumption involves stored energy, undigested food, and metabolic heat. Birds employ various adaptations to regulate heat and cut energy costs [5, 105-107]. The MEPR postulate for bird flight formation predicts the emergence of the most optimal self-organized pattern, the "V" formation, which is also the most commonly observed [105,106,107]. Additionally, the "V" formation allows for slight adjustments to accommodate birds of different sizes by providing enough lateral movement. When forming a pattern with the highest entropy generation rate or the highest rate per unit of energy, a migrating flock organizes into its most energy-efficient and stable configuration. The V and related formations enable spatial separation, helping maintain optimal bird temperature during flight. This formation also ensures that all trailing birds experience similar thermal environments. As flock size increases, the power expended per bird decreases when keeping the same velocity or mass flow rate. Compared to flying alone, flock formation reduces individual fatigue, allowing the flock to cover longer distances. The thermodynamic model supports three key experimental findings: (i) the preference for a V formation over other patterns, (ii) the energy-saving benefits of flying in a V formation, and (iii) the reduction in energy use as flock size grows (Figures 6a and 6b). The angle of the V depends on the birds' dimensions in the pattern, assuming uniform size across all birds. Table 2 provides size variation data for Canada Geese [105-107]. The model predicts a V angle of 113-116 degrees, calculated as 2 × tan^-1 (Wingspan/Length) for optimized flight formations. This prediction aligns well with reported angles for small flocks [105, 106, 107], though some variation is expected due to differences in goose sizes. The results suggest that formation flying, based on the thermodynamic model under MEPR conditions, has a strong physical basis in energy analyses, in addition to the behavioral explanations sometimes cited in ornithology [105]. It is worth noting that nearly forty-one bird species that undertake long migratory routes frequently fly in V formations [5], which may indicate that thermal herding is vital for flight efficiency. The V formation helps avoid creating new boundaries between birds while maximizing entropy production rate (Figure 6 (a) and (b)). This also means that the stored work or thermal energy gained from formation flying can be fully utilized for additional power when needed. As flock size increases, adjustments to the distribution of bird sizes may be necessary to maintain the formation. A wingspan distribution will alter leg separation but won't affect thermal calculations, as the model assumes energy transfer only along the flight path. A lower median length/wingspan ratio will result in a sharper V angle. This suggests that the MEPR-based prediction of 113°-116° is a reasonable estimate for larger Canada Geese flocks, which may display higher V angles. Larger angles are mainly observed in very small flocks [5, 105, 106], where behavioral and visual conformity might be prioritized over energy optimization, possibly due to easier coordination in smaller groups. Sharper angles may also occur if birds are not perfectly aligned in the direction of flight, as noted in Reference [107], or if a slight variation from the ideal V shape is chosen (several of which are discussed in Reference [5]). The V formation and similar formations could offer additional benefits beyond those provided by fluid-flow models that emerge from thermodynamic analyses. Without detailed force analysis, thermodynamic models can still evaluate the advantages of the formation. Tables 1 and 2 compare MEPR results with experimental data and contrast them with models based on synchronized flapping work. The maximization of entropy rate accounts for both equilibrium and steady-state behaviors, highlighting the ability of thermodynamic models to address pattern stability, whereas mechanical flapping-force models may not consider stability or the V-angle [5]. This does not mean flapping models are inaccurate; rather, it indicates they might lack thermodynamic optimization.

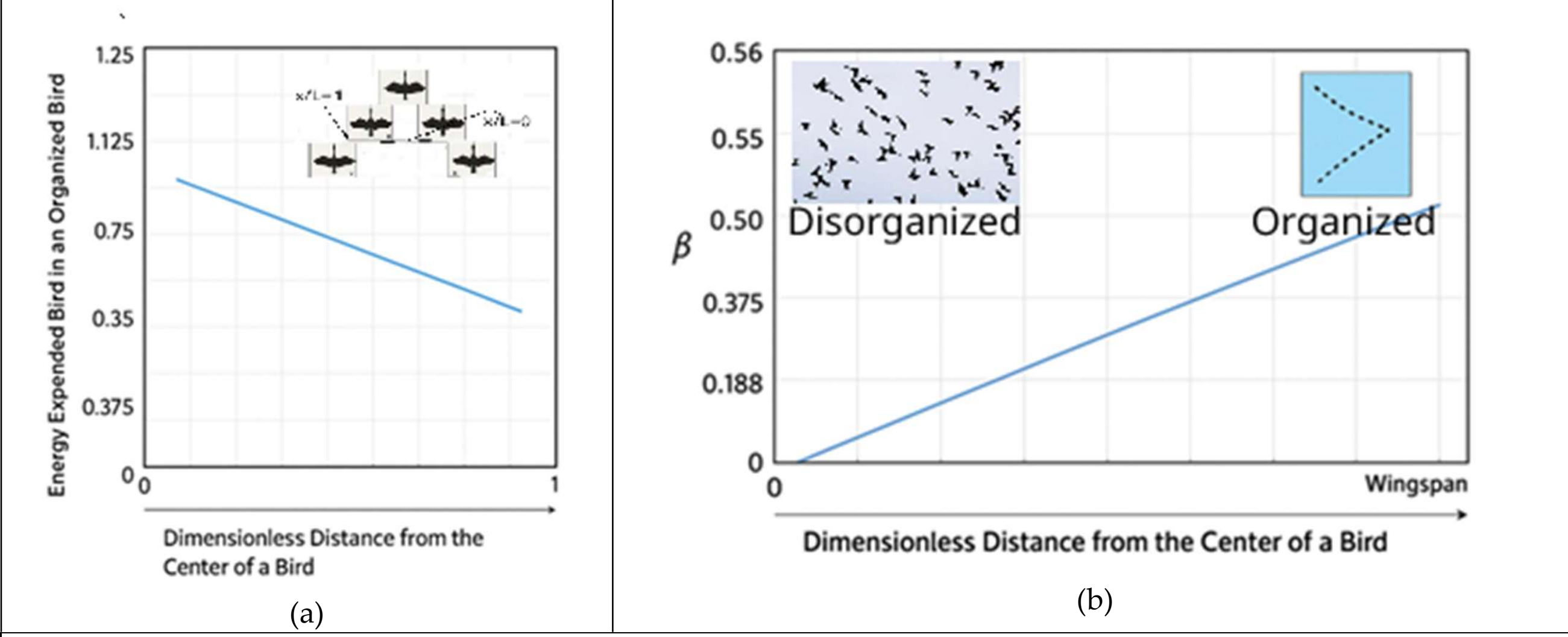

Figure 6. For the V patterns formed by birds in (a), i.e., when they self-organize into a V-shape pattern, they conserve energy per bird. The dimensionless distance ranges from 0 to 1 as shown in the inset [5]. (b) A plot β (K-1) (Entropy generation rate divided by the rate of energy expenditure per bird), as a function of the distance from the bird center, is shown for a three-bird flock [5]. The inset shows V formation from http://www.strategistblog.com/2020/10/flying-in-remote-formation.html and the disorganized state before forming into the V pattern for the long-duration flight.

Table 1. MEPR Validations across various types of studies. ml is the liquidus slope in the phase diagram, D is diffusivity, ∆T0 is the solidification range, k is the solute partition coefficient, Rg is the gas constant, and Tm is the melting point. G is the temperature gradient, and V is the solidification interface velocity

| Characteristics | Relation | Measurement range and MEPR assessment | MEPR validations and assessments. |
|---|---|---|---|
| Energy Conservation | Bird Angle for V formation flights of Canadian Geese [5]. | The observed V angle is ~ (110-130 degrees for a small flock of Canadian geese (05, 06, 107). See Table 2. | The angel for the V formation = 2.tan$^{-1}$ (Wingspan/Length). The median length/wingspan for Canadian Gesse, based on bird dimensions shown in Table 2 below, with MEPR analysis [5] is: 113-116 degrees. |
| Physical Constants | Diffusion Constants [62]. | **Pb-0.01 *wt*% Sn**<br>Experimental: 1.656 x 10$^{-9}$ m$^2$/s<br>From MEPR: (2.3 – 4.619) x 10$^{-9}$ m$^2$/s | **Pb-15 wt% % Sn**<br>Experimental: 1.656 x 10$^{-9}$ m$^2$/s<br>From MEPR: (1-67 – 73.54) x 10$^{-9}$ m$^2$/s (the lower value corresponds to a modified partition coefficient. |
| Bifurcations | Dilute Alloys. Critical tipping point predictions. | Plane Front to Non-faceted Perturbations. The symbols are defined in [4,23, 47, 48, 49,52, 62] $\frac{D_L}{\Delta T_O}\frac{2\, m_L\, \Delta h_{sl}}{T_m^2\, R_g\, \ln(1/k_{eff})} < \left(\frac{V}{G_{SLI}}\right)_C < \frac{2D_L}{\Delta T_O}\frac{2\, m_L\, \Delta h_{sl}}{T_m^2\, R_g\, \ln(1/k_{eff})}$ | Plane front to Faceted Perturbations. $\dot{\Phi}_{\max}$ *is the maximum entropy generation rate* $\dot{\Phi}_{\max} = \frac{\Delta\rho_k\, V^3}{2\, \zeta^2\, G_{SLI}}$<br>$\eta_G$, the number of atomic layers is equal to one, and the highest density planes are oriented for growth (4). $\zeta$ *is the diffuse interface thickness* |
| Climate Sciences | Upward velocity is measured for non-equilibrium thunderclouds that can produce rain [44] | Recorded measurements are about 10 m/s for cumulonimbus clouds (reported in [44]). | MEPR calculated: Depends on cloud thickness. Dor thicknesses between 1 km and 24 km, the predicted velocity is between 6 m/s to 11 km/s |

| Table 2: The Range of Physical Characteristics for Canada Geese [105,106] |
|---|
| Length: 29.9-43.3 in (0.76-1.10 m) |
| Wingspan: 50.0-66.9 in (1.27-1.70 m). |
| Weight: 105.8 - 317.5 oz (3-9 Kg) |

### 3.4 Analysis and Comparison of Entropy Generation and Morphological Patterns in Fluid Flows and Solidification

Continuous energy input sustains several stable patterns (morphologies) in the transformation of a system at a steady state in condensed matter, as discussed in the various examples in Section 1. The maximum entropy production rate hypothesis is the only solidification model to accurately predict bifurcations and high-temperature liquid diffusion constants across a wide range of solidification variables [4, 23, 62, 63], consistent with experimental data. MEPR boundary conditions model the energy needed to create and sustain morphological features as a work rate and include additional terms for the entropic contribution of defects. For patterns that exist in steady state due to external steady-state energy interactions, the morphological limits of typical patterns across a range of driving forces are often readily identified from experimental observations [47,48, 49, 59, 93, 89]. This section discusses known morphological limits on solidification and fluid flow under steady-state conditions from an entropy-generation perspective.

It should be noted that many morphological transformations remain poorly understood, although the limits and boundaries of specific patterns are generally recognized and shown in Figures 7(a) and (b). Transitions from one pattern type to another type happen at discrete points (called Bifurcation Points in Figure 7(a)). However, as noted in [4,47,48,49,52,54,62], there is often a region of bifurcation rather than a strict bifurcation point, which, in some cases, can lead to morphological metastability if the transition tipping point is not reached. The y-axis of Figure 7(b) is in units of J/(m3.K.s) for solidification and J/(m.K.s) for pipe flow, respectively, i.e., the plot of Equations 3.2, 3.3, and 3.6.

For steady laminar and turbulent flows, Pal [33] has derived entropy generation rates that depend on fluid properties, velocity, and pipe diameter, assuming constant temperature and negligible temperature dependence. Small eddies are isotropic, and their size is driven by viscosity. A critical velocity marks the transition between flow types (Figure 7(b), where energy is extracted by large, anisotropic eddies that are unaffected by viscosity [90,153,154,157]. A critical velocity or cooling rate exists where morphological transitions occur, with each morphology associated with a characteristic slope of $\dot{S}_{gen}$ versus velocity. For steady and fully developed laminar flow and for turbulent flows of a Newtonian liquid in a smooth pipe, the entropy generation rate **per unit length** of pipe is derived by Pal [33] as:

Laminar Flow:

$$\dot{S}_{gen} \sim (8\pi/T)\mu V^2 \tag{3.2}$$

Turbulent Flow:

$$\dot{S}_{gen} \sim (0.079\pi/2T)\mu^{0.25}\rho^{0.75}D^{0.75}V^{2.75} \tag{3.3}$$

The relationship between Reynolds number and entropy generation rate increases with flow speed, as shown in Figure 7(b), in a manner similar to bifurcations that are seen during solidification morphology evolution. Note that both solidification and fluid flow indicate slopes changing for the entropy generation rate vs driving forces (Equations 3.2 and 3.3). However, unlike solidification, pictorial descriptions of self-organized turbulent flow do not always present a single, universally accepted image [10, 91-95]; instead, they offer a conceptual view of turbulence development and organization within a fluid characterized by varying eddy sizes. Such diagrams display chaotic, swirling patterns of eddies and currents superimposed on the mean flow, in contrast to smooth, laminar flow that is readily distinguished from a turbulent morphology. Typical morphologies of both types are shown in Figure 7(a). The figure highlights only the more common types of boundaries that various morphologies can form [4, 33, 47, 48, 49, 86, 87, 88, 89]. Turbulence involves eddies of different sizes; at high Reynolds numbers, a scale separation exists between large and small eddies (an analogy here is to smooth cells, cellular dendrites, and fully developed dendrites). Turbulent flow is diffusive, with chaotic eddies enhancing boundary layer momentum transfer, delaying separation, increasing resistance and heat transfer, and dissipating kinetic energy into thermal energy. The largest eddies break up as the Reynolds number increases, transferring kinetic energy to smaller eddies. These processes continue until the eddy size is reduced to the smallest size and the eddy motion is stable with heat production from viscous effects. The energy dissipates as heat, and rotational inertial work from the overall kinetic energy of turbulent flow is thus converted into heat and boundary work, analogous to the potential energy (bond energy creation or lattice energy) during solidification. The morphology of eddies approaching and exceeding the energy input required by the flow at the Kolmogorov (small-eddy) limit remains poorly studied, but branching and optimized structures for energy dissipation are thought to be possible [90,153,157].

Note that Equations 3.2 and 3.3 are developed for steady state conditions. In laminar flow, entropy increases linearly with velocity; in turbulent flow, the increase is steeper and shifts upward with larger pipe diameters. Higher Reynolds numbers indicate greater turbulence, a broader eddy spectrum, and greater scale separation, with energy cascading to smaller scales where it dissipates (Equations 3.2-3.4. The smallest eddies that also form are particularly viscosity dependent and are described by Kolmogorov's limit [90,153,154] discussed below. In both turbulent and laminar flows, the entropy generation rate per unit length of the pipe depends on fluid density ($\rho$), as well as viscosity ($\mu$) and average fluid velocity (V). The temperature T is assumed to be constant in the analysis. Although this assumption isn't strictly valid in entropy generation analysis, Pal [33] suggests that the temperature dependency may be small for pipe flows. This approximation is invalid when eddies approach the Kolmogorov scale. Note the dependency on pipe diameter (D) for turbulent flows. The entropy generation density rate is always dependent on the pipe diameter. The smallest eddies have universal characteristics independent of the flow geometry and conditions. Those eddies in this range usually receive energy from the larger eddies and dissipate their energy into heat through the fluid's molecular viscosity. These eddies are isotropic, with length scales characterized by Kolmogorov scales [157]. It is assumed that the small-scale eddies are determined by viscosity, and the specific dissipation rate density of the turbulent kinetic energy is ε. The wave number κ (for eddy size $L=2\pi/\kappa$) is related to the Reynolds number. The energy $E(\kappa) = C_k \kappa^{-5/3} \varepsilon^{2/3}$ is the energy spectrum function (units of $m^3/s^2$ or J.m/kg), which changes with eddy size or curvature and $C_k \sim 1.5$ [94]. A high Reynolds number is required for morphological scale separation to become easily visible [157]. The ratio of the largest to the smallest eddy scale (η) scales with the Reynolds number [90] and is given in the following manner:

$$\frac{L}{\eta} \propto \mathrm{Re}^{3/4} \tag{3.4}$$

Re is the Reynolds number for a Newtonian flow, and L is the largest eddy. Consequently, the morphological transitions are related to the $S_{gen}$ rate in turbulent flows via Equations (3.3) and (3.4). The heat-transfer rate per unit wall temperature is proportional to the entropy generation rate. Multiplying the entropy generation rate by the average temperature and differentiating with respect to a spatial coordinate can indicate the shear. Shear serves a dual role: it is a source of turbulence in the system (e.g., in boundary-layer turbulence), and it also alters energy transfer from large to small scales by affecting the Kolmogorov limits [90].

There are notable similarities between cellular-to-dendritic transitions during solidification and laminar-to-turbulent flows, as discussed below. Morphological limits to boundaries between facets, half cells, stubby cells, cells, cellular dendrites, dendrites, and low and high velocity plane front limits [4,47, 48, 49, 89, 92, 94, 104, 131], which will be referred to as the Trivedi limits [59, 94, 156]. The cooling rate during solidification in a positive temperature gradient environment (i.e., dT/dx>0) is similarly related to the entropy generation rate [4, 23], thereby indicating that the Trivedi limits on morphological boundaries depend on it. Likewise, there is a threshold beyond which cell and dendrite boundaries cease to form, resulting in a planar interface (most likely with a highly diffuse interface) at very high solidification velocities. Mullins and Sekerka first described the formation of high-velocity planar interfaces [47, 48, 49]. This was later recognized as possibly a very wide, diffuse interface [4,23,52,93,94,104]. The transition to a glassy, non-crystalline phase is also attributed to the high-velocity transition, and the associated non-crystalline phase to the entropy generation rate [17, 18, 52, 163]. Dissipation in a liquid depends on the square of the velocity gradients (how sharply the speed changes over a tiny distance). Even if the viscosity is small, the "shearing" between turbulent eddies becomes so intense at high speeds that the total energy converted to heat increases significantly. Regardless, it must be remembered that at high Reynolds numbers the coefficient of friction falls only very marginally with the Reynold number. The total drag force is proportional to $V^2$. The energy dissipation is proportional to $V^3$. This assumes that the production of eddies matches the destruction of eddies. The rate of internal work to form eddies scales with $V^3$. The energy flux moves down the scales. For fluids, this is also the dissipation rate.

There are examples in which heat is converted back into turbulence. The sun heats the air (Entropy/Internal Energy), which creates the massive turbulent eddies we call wind and storms. This was dealt with in Section 3.2. More commonly, when the Richardson number (Buoyancy Production/ Shear Production) is negative, the heat actively creates eddies by doing work on the system. The heat adds energy to the turbulent kinetic energy. This creates a "thermal" or a "plume" described in reference [99] (see appendix in the article).

Both the Kolmogorov-scale eddy development for fluid discussed above and the tip of a dendrite during solidification impose limits on morphology within specific domains of entropy generation rate. During solidification, the Trivedi limits apply for the morphological bounds of the dendritic tip radius defined by the relationship shown in Equation (3.5). During viscous dissipation, heat is produced. The eddy size variation dependency on the Reynolds Number is like the known variations in tip radius size of dendrites (and the secondary arm spacing) with the Peclet number during solidification (Equation 3.5) [47,48, 49, 88, 89, 92, 93, 94], but as pointed out below, also more related to the secondary dendrite curvature distribution.

$$R_{tip}/R_{tip\text{-}min} = Pe \cdot (V_a/V) \tag{3.5}$$

Where Pe is the Peclet number, $V_a$ is the absolute velocity at which dissipation reaches its maximum (i.e., boundaries disappear and a broad diffuse phase is expected to form) [4,47,48,49,52,92,93,104], like the Kolmogorov limit [90,157]. From references [23, 62], one can relate the cooling rate to the $\dot{S}_{gen}$ (entropy generation rate J/K.s) at the critical bifurcation point to the cooling rate (V.G) by an expression shown in Equation (3.6), where V is the solidification velocity, G is the temperature gradient, and $\Delta\rho_k$ is the normalized density difference between the transforming phases. The subscript SLI refers to the solid-liquid interface region, and c indicates critical. Nc is a constant that includes solute diffusivity. This reduces to ($2\Delta hm/\Delta\rho_k Tm^2$) [62] for pure metals.

$$|(VG_{SLI})_C| = |\dot{T}| = \frac{2\,(S\dot{gen})_C}{\Delta\rho_K\,\mathbf{N_C}} \tag{3.6}$$

Figure 7(a) highlights the type of boundaries various morphologies can form [4, 47, 48, 49]. Equation (3.6) indicates that the $\dot{S}_{gen}$ scales with the cooling rate, as shown in Figure (7b). Note that, as discussed in reference [4], the $\dot{S}_{gen}$ is associated with simultaneous boundary creation (stored work).

The entropy generation rate and the solidification rate are related as shown in Equation 3.6 [23], which can be formulated as the diffuse interface size variation with the cooling rate during solidification [62], shown in Figure 7b, which is a schematic of the behavior of the specific entropy generation rate with increased input power that can cause morphological transformations in condensed matter [23, 33, 62]. A comparison of boundary spacing between experimental measurements and model predictions indicates that $\dot{S}_{gen}$ is maximized even when solute temperature gradients (another form of entropy generation condition) are considered [23]. The crystallization case has more experimental studies related to the Trivedi limits [47,48,49,89,93,94], thereby providing more detailed information on the morphological limits during solidification. The Kolmogorov limit suggests that energy fluxes dissipate towards small scales, eventually leading to the significant dissipation of kinetic energy [90,91,92,157]. However, for fluid flow, an estimate of energy dissipation is only made with difficulty because the energy dissipation occurs with unclear boundary formation information (diffuse), albeit with some spread in curvature, like the spread in dendrite arm sizes shown for secondary dendrite arms in Figure 7.

In this section, the two examples show that thermal energy and sub-boundary work influence morphological and pattern stability and set morphological boundaries for systems that support pattern formation in condensed matter. Although Equations 3.2, 3.3, and 3.6 are steady-state formulations, the second derivative of the entropy generation with respect to time is negative whenever studied [23, 62], indicating a maximization as long as the first derivative is zero (inferred from the linear relationship). The significant role of thermal effects in fine-scale morphological transitions is evident in all experimental studies at the Kolmogorov limit [90] and in the high-velocity Trivedi limits [89,94], which are approached by a transition to finer cellular structures, followed by a very wide, diffuse interface or a phase transition.

### 3.5 Entropy generation in solid-state transformations like sintering, with re-arrangement of sub-boundaries.

All self-driven sintering processes produce entropy ($S\dot{gen}$) [6, 14, 26, 31, 47, 50, 51, 62, 64, 95, 96,97]. Entropy generation arises from heat conduction, mass transfer, reactions, and grain motion. Rapid sintering releases enthalpy: configurational entropy influences reactions and grain-boundary migration, thereby affecting properties such as hardness and toughness. Slowing sintering can prevent catalyst coalescence, allowing lower-cost high-temperature options. The ratio of grain boundary to surface energy ($\gamma_{gb}/\gamma_s$) governs sintering, requiring that grain-boundary formation be energetically favorable. Additives and pressure assist the process. Grain growth decreases the total boundary area, thereby lowering the free energy. Particle sintering involves heating or temperature gradients that cause grain-boundary expansion and porosity reduction, driven by differences in chemical potential and interfacial energy. Convex areas have higher chemical potential, thereby diffusing atoms to lower-potential regions, reducing surface area, and releasing energy. Sintering pathways occur with or without grain growth, depending on whether surface and grain-boundary energies are similar; densification without grain growth occurs if grain-boundary energy is much lower than surface energy; often, both occur simultaneously, starting with pore elimination. Kinetic factors influence neck formation at grain contacts. Whether MEPR applies to sintering is unconfirmed; future research is needed. Sintering likely follows a maximum-entropy production pattern, consistent with S-curve behavior discussed in Sections 4 and 5.

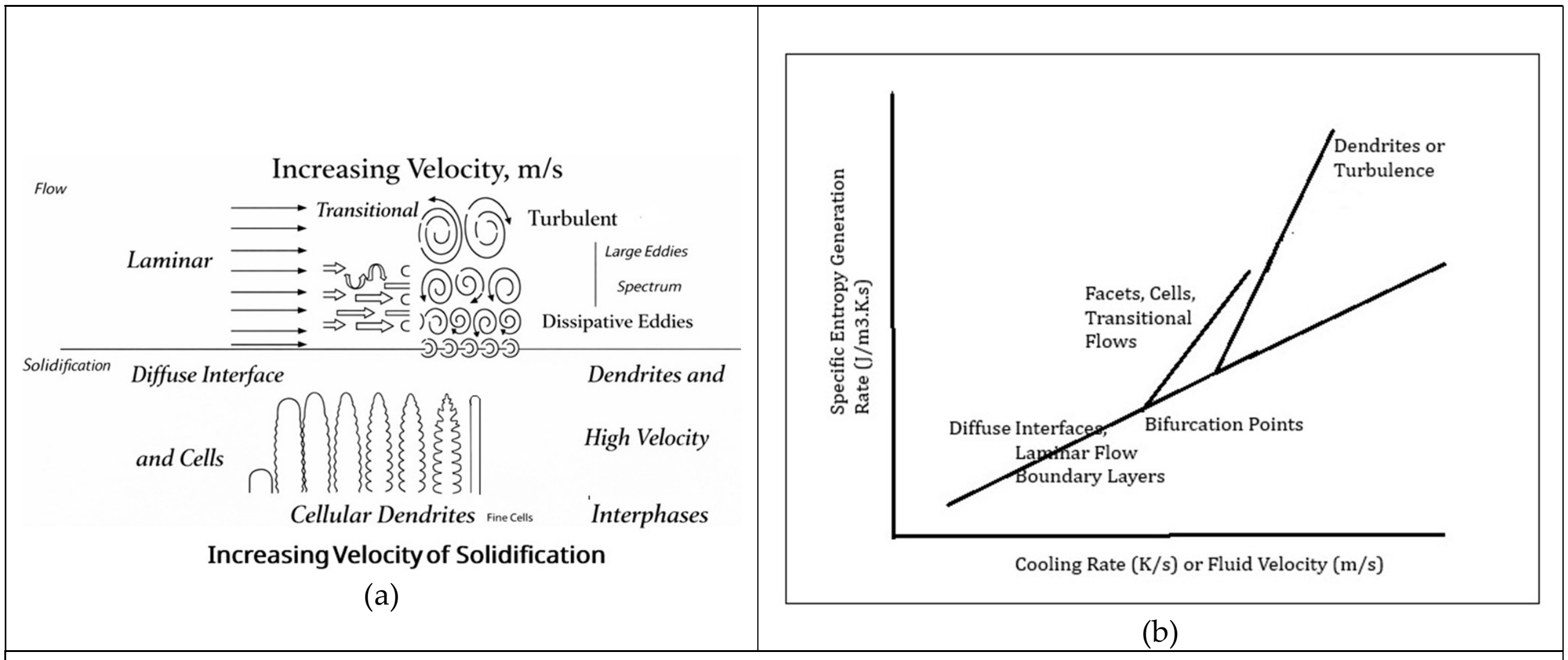

Figure 7. Schematic illustrations of laminar, transitional and turbulent flows and cellular and cellular-dendrites or dendrites, (b) The rate of entropy generation increases with the driving force (rate of energy exchange) from an change in the solidification velocity from reference [62] or fluid velocity in a circular pipe of diameter D from reference [33], i.e., morphological change with an increase with the driving forc. respectively. Both axes are plotted on a log scale. A change in slope indicates the emergence of a new morphology at a bifurcation point. The y-axis of Figure 7(b) is in units of $J/(m^3.K.s)$ for solidification, and J/(m.K.s) for pipe flow, respectively, from Equations 3.2, 3.3, and 3.6.

# 4.0 Ease of Computation with the MEPR Condition for Steady State and Non-Steady State Self-organization.

The MEPR condition discussed above permits kinetic determinations, particularly when morphological transitions are involved. As we will note later, these are important for understanding the stability and resilience of complex structures. Two examples are provided below: one from a steady-state, one-dimensional morphological transition and another from a non-steady-state, nucleation-triggered transformation that leads to an ordered self-organizational process, offering further validation and extensions for understanding the significance of the rate of entropy generation as a possible principle applicable to several aspects of self-organization.

## 4.1. Morphological Transitions Modeled Under Steady-State Conditions.

Dendritic patterns resemble tree roots and occur in nature to facilitate efficient transport and distribution. Blood vessels and nerve cells use dendritic structures to optimize oxygen flow and signal transmission. River networks and plant roots spread in dendritic patterns to distribute water and nutrients. The solidification analogs developed by Trivedi et al. [47, 48, 49, 54, 59, 94] have greatly improved the understanding of such systems. Yet there is no clear thermodynamic basis for morphological transitions rooted in a firm thermodynamic approach, except for steady-state solutions described in [4, 6, 23, 29, 39, 62, 63, 104]. In this article, the analysis includes non-steady behavior. Predictive morphological models based on the MEPR principle can predict features such as changes in dendrite arm spacing during solidification in response to process variables, such as solidification velocity (V) or spacing, by setting $\delta/\delta\lambda(\delta Sgen/dt) = 0$ or $\delta/\delta V(\delta Sgen/dt) = 0$. This condition helps approximate the connections between morphological and connectivity features within a self-organized system or subsystem. The following equation from reference [52] captures the energy cost of maintaining the morphological features that export entropy from a control volume, which exports boundary entropy and a new boundary volume (including subboundaries) with higher energy than the lattice. When a morphological transformation occurs, $\lambda 2$ (secondary dendrite arm size) responds to new defect evolution conditions during a self-organization process. Following the arguments in reference [52], when a transition to dendritic morphology happens, it occurs from a cellular dendrite morphology to a dendritic morphology.

$$(\lambda_1^2/\lambda_{2(C-D)}) = 6\boldsymbol{\Gamma}\cdot(T_s/T_l)/(\Delta T_0 - \Delta T_{ctip(C-D)}) \quad (4.1)$$

$\Delta T_0$ is the solidification range, and $\Delta T_{ctip(C-D)}$ is the difference in the tip temperatures between the two morphologies. Here, Ts and Tl are the solidus and liquidus temperatures, C-D indicates cellular to dendritic, and **Γ** (the capillarity constant) is typically about 10^ (- 7) for metals [47-49]. Figure (8) is a plot of $\lambda_1$ vs. $\lambda_2$, for metals from equation (4.1) recast as Equation (4.2) with the clubbing of material constants in a catchall constant C.

$$\lambda_1^2 = C.\lambda_{2(C-D)} \quad (4.2)$$

Note that Figure 8 has been plotted solely from the thermodynamic constants in Equation (4.2), yet it connects various emergent scales in the solidification microstructure. The $\lambda_1$ at the C-D transition condition is reported to be ~300 micrometers for two alloys with very different solidification ranges $\Delta T_0$ (369 K for the Rene-108 alloy, but only about ~29 K for the IN-718 alloy). Figure 8 is predictive for both alloys.

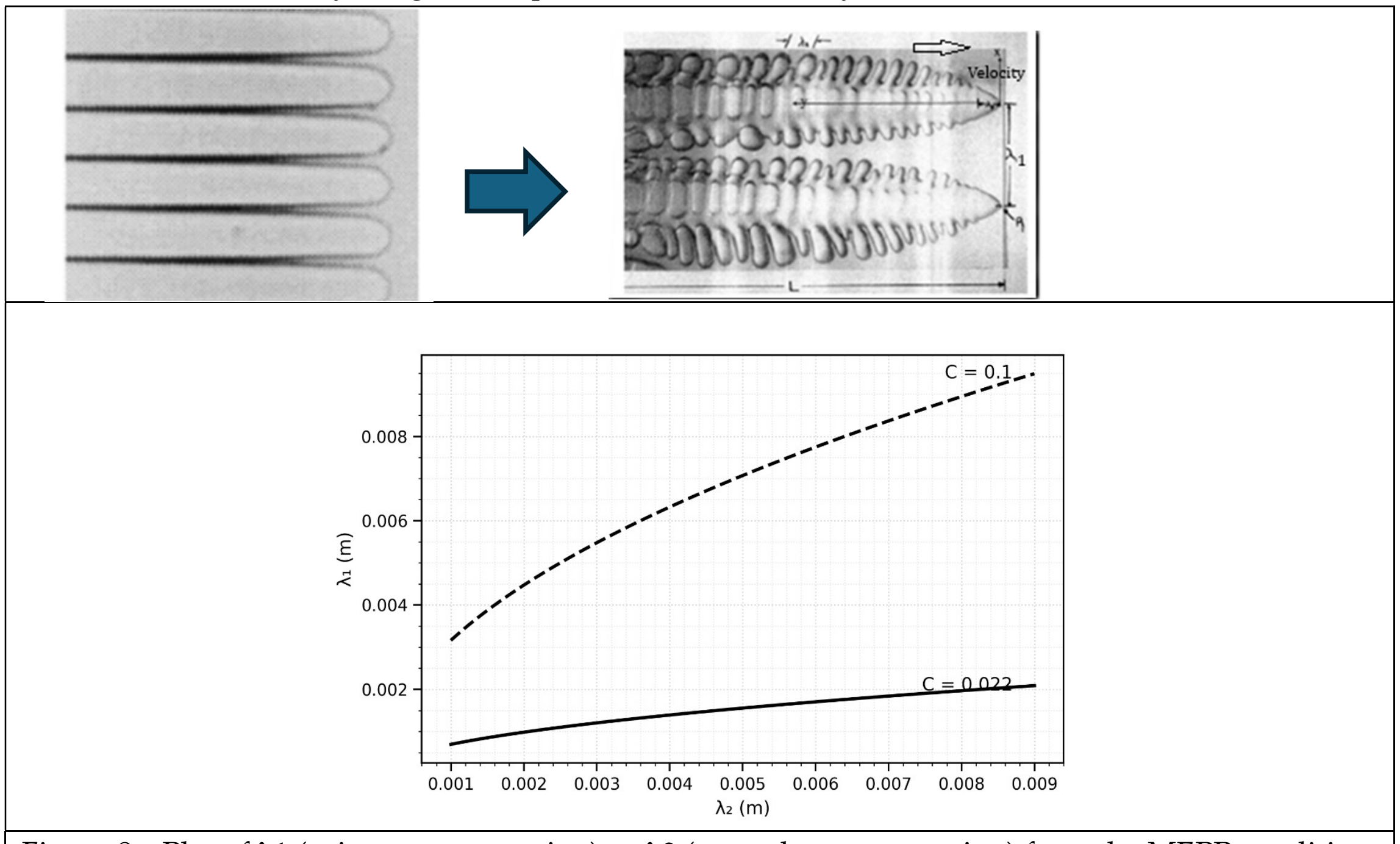

Figure 8. Plot of λ1 (primary arm spacing) vs λ2 (secondary arm spacing) from the MEPR condition. Dendrite figure with λ1 (primary arm spacing) vs λ2 from reference [89] used with permission.

Note the small to large secondary arm dendrite variations of the secondary dendrite arm distance (or curvature) from tip to root, which mimics the spread in eddy sizes described above in the previous section. Although not studied further in the article, the dependence on the Peclet and Reynolds numbers is noted to be similar.

### 4.2 Entropy generation for time-dependent or non-steady-state cooling.

As an example of a solution for a non-steady state solidification problem, consider the temperature or pressure-based undercooling of a volume of liquid (like a small droplet) that occurs before a recalescence event [46-50, 75, 76]. Assume a small droplet with a low Biot number is cooled into a supercooled state (assume that the droplet is a sphere with radius R, h is the heat transfer coefficient, and $\kappa$ is the thermal conductivity (assumed equal for the solid and liquid). The droplet's temperature-time profile is typically of the type shown in Figure 3. The cooling can be assessed in several parts. The first part occurs during droplet cooling, without any recalescence. The second part involves recalescence, in which the fraction solid transitions from 0 to 1 over a very brief period (this part is quasi-adiabatic if the transformation is rapid); the third part is the solid's cooling. In the first and third parts, work generated during entropy generation is due to compression work (i.e., because the rate of thermal expansion with temperature is positive for all materials). Cp is the liquid specific heat, R is the droplet radius, T is the temperature in Kelvin, $T_0$ is the ambient temperature, and h is the boundary heat transfer coefficient. Small Biot number conditions are assumed for simplicity, although it should be recognized that temperature gradients in the recalescing solid interface into the liquid will also produce entropy [23,62]. The cooling in the liquid or solid droplet states with the MEPR principle gives:

$$\mathrm{d}(\mathrm{d}Sgen)/\mathrm{dt} = \mathrm{Cpl}[(\mathrm{T}^{-1}.\,\mathrm{d2T/dt2}) - (\mathrm{T}^{-2}(\mathrm{dT/dt2})] + [\mathrm{h/R.\,T0.\,T^{-2}dT/dT}] + [\mathrm{PT} - \mathrm{2dT/dt}] = 0 \qquad (4.3)$$

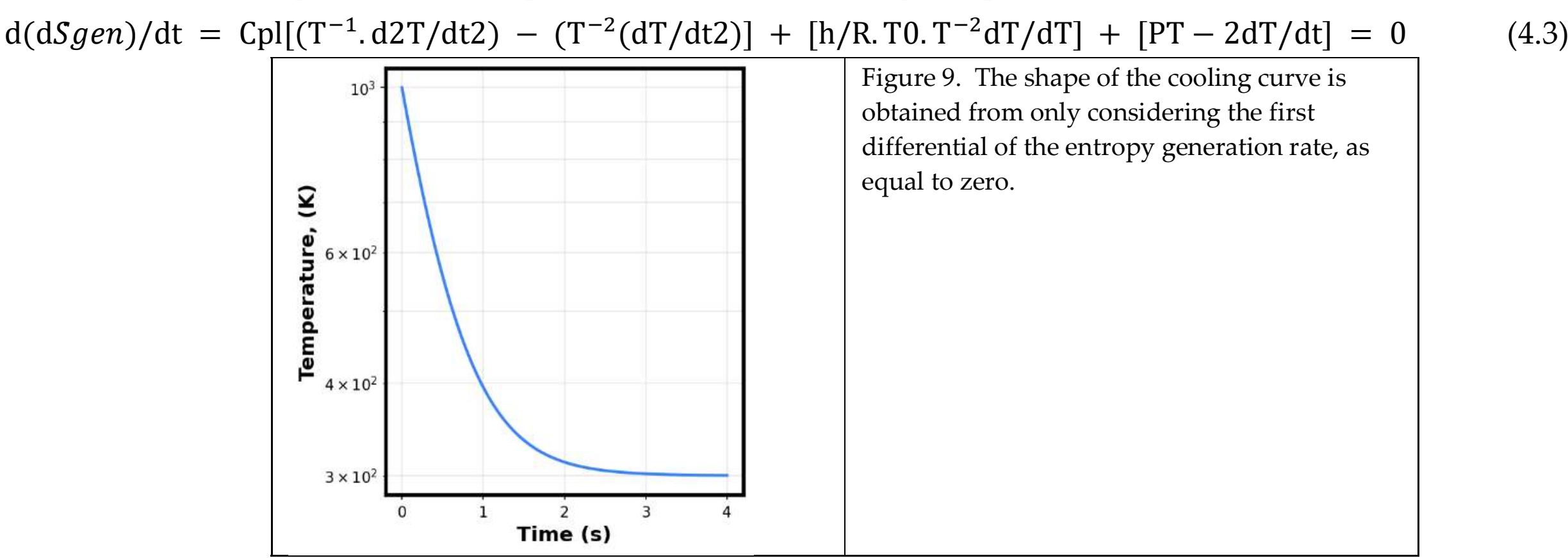

Figure 9. The shape of the cooling curve is obtained from only considering the first differential of the entropy generation rate, as equal to zero.

Assuming Cp =1000J/K.s, h=$10^5$ J/m$^2$K.s, R=$10^{-3}$ m, $T_0$=3x $10^{-2}$K.,the compressive pressure P=10 J/m$^3$ equation 4.3 can be solved with the initial conditions T(0) =1000K, dT/dt (0) =$10^3$ K/s. Figure 9 shows the shape of the cooling curve.

The plot (logarithmic y-axis) in Figure 9 shows that, $T(t)$decreases exponentially due to the damping term $\left(\frac{h}{R}\right) T_0 term$.

In Equation 4.3. The shape of the cooling curve is the expected solution for all cooling processes and thus validates the conditions developed in Section 2 regarding the entropy generation rate: the process path maximizes the entropy generation rate, i.e., the time derivative of the $S_{gen}$ rate is zero. For completion, it should be noted that if this curve were a representation of a statistical function of the form, $f(T) = \lambda e^{-\lambda T}$, where $\lambda$ is a rate parameter; the entropy and entropy generation rates are positive and always at a maximum [82, 83, 84]. Such an analysis applies to the information-loss rate of Shannon entropy. In the sections that follow, emphasis is placed on S-curve behavior. It should be noted here that solutions that are exponential can be considered a part of an S-shaped curve. The role of $\lambda$ is that it suggests an exponential decay curve if ($\lambda$>0), an exponential growth curve if ($\lambda < 0$), and a horizontal line if ($\lambda$=0). It is also known that among all continuous probability distributions of a variable between (0, ∞) that have a mean $\mu$, the exponential distribution with $\lambda = 1/\mu$ has the largest differential entropy [19,34] (see also Appendix A). In other words, it is the maximum entropy probability distribution for a random variation, which is greater than or equal to zero, and for which the expected value is fixed. This is yet another indication that the maximum rate of entropy production, which is associated with stored work and energy, supports the confinement process required for the distribution. Appendix A, along with Sections 4 and 5, discusses the entropy generation rate for distributions with no skew (Gaussian) and for distributions with skew (Exponential).

For the recalescence part of the droplet cooling/solidification process [46-49, 75, 76], a MEPR condition that ignores external cooling for a linear droplet [76] yields a second-order differential equation of the type, which in its simplest form can be written as

$$\frac{d2l\ \left(\frac{T}{T0}\right)}{dt^2} = -\left\{\frac{C.Tm}{T.L}\right\}\left[\left(\frac{Tm-T}{Tm-70}\right)^n\right]\left(\frac{dln\left(\frac{T}{T0}\right)}{dt}\right) \tag{4.4}$$

Here, $t''$ is the dimensionless time ($D.t_r/L^2$), where t(s) is real time, D is diffusivity (m2/s), L is the radial dimension, L is the scale of the problem (e.g., the droplet radius~$10^{-4}$m)., fs (T) is the solid fraction, T is the temperature, Tm is the equilibrium melting point, and $T_0$ is the temperature of the reservoir, outside the physical boundary of the cooling or recalescing droplet. C is a constant $\Delta hsl$.Kc/$C_p$, where $\Delta hsl$ and $C_p$ are the heat of fusion and specific heat per unit volume, and Kc is the linear interface kinetic constant ($m.s^{-1}.K^{-1}$). C has units of velocity, namely m/s, which indicates an average transformation rate for a transformation of this type. In Equation (4.4), the term raised to the n-th power is used to account for curvature. In Equation 4.4, the term in the bracket raised to the power n compensates for curvature and other overall geometric shapes of the solid liquid interface

Equation (4.4) is solved for a metal droplet with properties resembling an aluminum alloy, $\Delta hsl$= 1.6. $10^9$(J/m3) (high end) and Cp= 2.43. $10^6$ J/$m^3$.K, and Kc=0.01 m/K.s, or similar, using the conditions T(0) = $T_N$= 700K, the assumed nucleation temperature, and the interface velocity condition described in the caption of Figure 10. Specifically, the velocity is proportional to Kc.(Tm-T), where Tm is the melting point (933 K for aluminum). The solutions for T(t) and dT(t)/dt are shown in Figure 10(a) for C/R= 4.45 x 10^5 (R=140 micrometers) and indicate a very limited effect of n (tested from 0.8 to 1.6). All the solutions discussed below are for n=1. The solution (Figure 10(a)) indicates an extremely high initial heating rate. Note that the temperature plot is sigmoidal, whereas the solution indicates an initially very high heating rate of approximately 10^8 K/s that subsequently decreases with time. The initial heating rate, however, is approximately an order of magnitude higher than in typical experimental reports [165] but shows similarity to other numerical calculations [76]. Note that initially the interface could be considerably high in the radius of curvature, which will damp the initial temperature rise – a feature not considered in Equation 4.4.

The sigmoidal shape, as shown in this section, is a characteristic of a normal distribution, as indicated by its PDF (Probability Density Function) and CDF (Cumulative Distribution Function) (the sigmoidal curve itself), and is a key feature of self-organization in complex systems. Appendix A shows that any sigmoidal process can be treated as the PDF and CDF functions of standard statistical distributions, which can then be used to infer the entropy generation. The PDF and CDF will define ratios analogous to the entropy generation rate and will affect the boundary formation rates. The rate of entropy generation, which is the ratio of the PDF to the CDF, is maximized at every point on the curve [61], but in some cases reaches its peak at the beginning and stays constant. A steeper CDF curve for a normal distribution (PDF) indicates that data points are more tightly grouped around the mean during the self-organizing process, meaning less spread, a smaller standard deviation, and lower variance. Since the PDF of a normal distribution is identical to the distribution itself, several observations can be made that provide more general insights into self-organizing processes by noting the following:

(a) The PDF/$(CDF)^2$ can approximate the entropy generation rate as shown in Figure A3 by the cumulative integral.

(b) The cumulative integral of PDF/$(CDF)^2$ ratio is always positive, as is required for the entropy generation Sgen and $\dot{S}_{gen}$ (rate).

(c) An asymptote is noted at infinity ($z \to -\infty$ *or the start of the sigmoidal start plateu*), because the CDF approaches 0, and the PDF also approaches 0, but the ratio approaches a very large, positive value.

(d) Note from Appendix A that the PDF/$(CDF)^2$ for both normal and skewed distributions shows little variation beyond z=2.

(e) For the standard or normal distribution: At the mean ($\mu = 0$), The PDF is at its peak. The PDF/(CDF)2 quickly falls to near zero for both the normal and exponential distributions. There is a relatively stable point of entropy generation across the central part of any such processes..

(f) As the PDF/CDF ratio ($z \rightarrow \infty$ $i.e. the\ begining\ of\ the\ high\ plateau$), the normalized CDF approaches 1, while the PDF approaches 0. The ratio, therefore, approaches 0.

(g) For the graphs shown in the Appendix, for the cumulative integral approach 1 for the Gaussian and 0.59 for the Exponential. This is the maximum stored for integration from z=0 to z>2.

The solution to Equation (4.4), which can be written in statistical distribution terminology as *F(T/To),* is a sigmoidal function. The solution to Equation (4.4) resembles the CDF of a normal distribution, i.e., a Gaussian function, Equation (4.5). The heating or cooling rate is obtained from the Gaussian (PDF). These are detailed in Appendix 1, in Equations A1 and A2, with the solution to the cumulative provided in Figure A3. These statistical function analogs represent the functions defined in equations 4.5-4.9. Consequently, the temperature, the rate of change of temperature, and the entropy generation for zero skew and exponential skew can be determined. Since the error function approximates the CDF of a Gaussian, an error-function fit can also be applied to the curve, as illustrated in equation (4.6) and in the generalized analyses in Figures A1 and A2 (Appendix A), using the boundary conditions for the recalescence problem discussed above. In equation (4.5) $\mu$ and $\sigma$ are the mean and standard deviation, respectively, which, for sigmoidal-shaped curves, are illustrated in Section 5. The PDF and CDF (S-shaped curves) for each distribution are shown in Figure A1.

$$f(T/To) = \dot{\mathrm{T}}\ /\mathrm{To} = d/dt\left(\frac{1}{\sqrt{2\pi\sigma^2}}e^{-\frac{(T/T0-\mu)^2}{2\sigma^2}}\right) \tag{4.5}$$

The rate of change of temperature is the derivative of *F(T/To) with respect to T/To, i.e., f(T/T0), which can be written as* shown in Equation (4.9), which is the PDF of the error functions as shown in Appendix A. Because the T(t) plot resembles an error function, curve-fitting methods can be applied to the parameters describing Figure 10(a). To simulate the key aspects of a typical S-curve solution, assume that the curve begins at T(0) =$T_N$= 700K, and levels off at T=933K. $T_N$ denotes the nucleation temperature, more generally, the tipping-point activity variable. The error function (erf) and its complement (erfc) appear naturally in transient, one-dimensional conduction when a boundary temperature or heat flux is altered (semi-infinite domain). It also reflects the Gaussian similarity structure of the diffusion operator and is ubiquitous in transport phenomena [92]. A set of fitted parameters for the error function approximation of the graph shown in Figure 10, that equilibrates between 925K and933K with a logistical fit corresponding to $T(t)=700+225/[1+\exp(k(t-t_0)]$ when k= -2.6$e^{-5}$ and $t_0$= 2.1.$e^{-6}$ are the logistic fit parameters. This reproduces the graph with an error function like fit of approximately $\mathrm{T(t) = a\ erf\left(b\ (t - c)\right) + d}$, as long as the constrain is imposed of $\mathrm{T(0) =}$ 700Kwith a=$2.187\times10^8$, b=$1.079\times10^5$ s−1, c=$-3.238\times10^{-5}$ s, d=(925−a). Thus, an error function fit, which is a well-known fit for an S-curve behavior, is noted for the solution. These parameters shape the curve to match the nonlinear recalescence behavior. The initial slope is very large, consistent with the order of the initial rise in temperature observed in previously published recalescence and rapid transformation results [64, 75, 76, 86, 138, 165]. Equation (4.4), plotted in Figures 10(a) and (b), indicates the recalescence behavior to be sigmoidal shaped. Sigmoidal solidification has also been discussed in [117], thus indicating that various mathematical descriptions of the problem, including the use of MEPR, yield identical solutions. Note that the sigmoidal curve is best plotted as the transformation or function of interest (like temperature) against non-dimensional time ($D.tr/R^2$) where R is the scale of the event. When discussing the sigmoidal curve below, we will assume that the functions plotted are always intensive functions on the y-axis (i.e., ones that do not depend on the size of the control volume) against non-dimensional time. This approach involves consolidating several similar plots from events of interest (such as densification, oriented boundary areas, or combinations) to facilitate a more comprehensive assessment of their behavior. When defined by the error function, the temperature and the heating rate plot shown in Figure10 are analogous to the CDF and PDF of standard distributions (see Appendix A). The PDF has the form:

$$\frac{d^2T}{dt^2}(t) \; = \; -\frac{2a\,b}{\sqrt{\pi}} \; \exp\left(-[\,b(t-c)\,]^2\right) \qquad (4.6)$$

In actual situations where the interface is advancing rapidly, temperature gradients [76] and sub-boundaries [113] are observed or calculated during recalescence, even when the dimensions of the region of study (e.g., the atomized droplet) are very small. For a moving, transforming interface, diffuseness at the interface and defect boundaries will be encountered, and entropy will be generated during the process. The entropy generated, $\dot{s}_{gen}$ (per unit volume), exhibits a behavior proportional to $\dot{T}/T^2$ [23, 52, 62], shown as an approximation in Equation (4.7). Entropy generation is also obtained by an entropy balance with a diffuse interface [4,6, 16, 18, 23, 52]. The heat of fusion of the solid with defects is, $\Delta hm$ ($J\ m^{-3}$), and the equilibrium heat of fusion $\Delta hsl$ ($J\ m^{-3}$).

$$\dot{s}gen \; = \; \left|\Delta hm.\left(\frac{\dot{T}}{T2}\right)\right| \qquad (4.7)$$

Equation 4.7 can be reformulated to yield the internal-boundary formation work, as shown in [4, 62], using Equation 4.8 and the MEPR ($d(\dot{s}gen)/dt = 0$) criterion to yield Equation 4.9.

$$\Delta hsl = \Delta hm + \omega_D \qquad (4.8)$$

where $\omega_D$ ($J\ m^{-3}$) is the energy of defects (such as grain boundaries or dislocations) per unit volume. $A_D$ is the area of subboundaries (per unit volume), and $dA_D/dt$ is the rate of formation of subboundaries (per unit volume). If the $\omega_D$ term is considered in the entropy balance, then the rate of $\dot{\omega}_D$, namely, $\gamma.dA_D/dt$, i.e., related to the rate of new boundary formation rate $dA_D/dt$ ($m^2/s$) (per unit volume), which has the units for diffusivity and $\gamma$ (the grain boundary energy per unit area ($J/m^2$). The logistic $T(t)$ fit given above, which matches the ODE-generated thermal curve in Figure 10, is used to solve Equation 4.9.

$$\frac{dA}{dt} = (\Delta hsl - \gamma.A_D)\ddot{T}(t)/\,(\gamma.\dot{T}(t)) \qquad (4.9)$$

Figure 11 (a), (b), and (c) are plots of $A_D$ and $\frac{dA}{dt}$, with the initial conditions $A_D(0) = 0$, to solve equation 4.10. The amount of boundary area per unit volume that finally forms is dependent on the $\gamma$($J/m^2$) as shown in Figure 11. Note that, as expected, the area of defects increases with the entropy generation rate, but starts decreasing at some time when $\left(\frac{\dot{T}}{T2}\right)$ decrease with time and become negative (indicating boundary elimination) when the second derivative of temperature with respect to time becomes negative, as can be seen from Equation 4.9. Note that the $A_D$ and $dA_D/dt$ are related by Equation 4.9 to CDF- and PDF-shaped distributions that reflect the temperature and rate of temperature change. The maximum $A_D$ timing is close to the logistic inflection time (~2.0 μs) and is only weakly dependent on the $\gamma$. That's expected because $\ddot{T}$changes sign at the logistic midpoint; since $\dot{T} > 0$, switches from positive to negative there. The entropy generation rate per unit volume, however, continues to increase as per Equation 4.7, and the corresponding analog equations shown in the Appendix (see Figure A3) continue to hold up to a point at which the temperature derivative with respect to time reaches zero. The implications of Gaussian and Skewed distributions may relate to the types of defect boundaries formed (thermal vs. athermal) and to external impacts on the control volume, but such an analysis is left to future studies.

Only a few experimental studies can validate the calculations because of the rapid changes in the variables involved; therefore, at this point, only the trend line of the predictions can provide some validation. Figure 12(a) from reference [113] shows a TEM bright-field image of Al-Zn particles aged at 130 °C for 1 hour after an atomizing process (for extremely small, atomized particles that fit the solution to Equation 4.4). A subsequent aging process reveals the boundaries as well as the two types of transformations noted in the particle, namely precipitates (or a metastable phase) labeled A and the serrated structure labeled B (picture, white lettering): note that a dark region that encompasses the entire particle for very tiny (50 nm), atomized particles. In the center of the photomicrograph is an extremely tiny pure Zn particle (~30nm), indicating the possibility of a highly supersaturated nonequilibrium phase formation during the early stages of recalescence. Note the facets in this particle, which are an indication that it is a product of the subsequent heat treatment. Figure 12(a) [113] depicts a scenario in which an atomized droplet begins to solidify after undercooling, and the interface velocity then decreases over time as indicated for the Kc considered for Equation 4.4 and plotted in Figure 10. Although not fully clear due to the fineness and a lack of understanding of the precise nucleation regions, Figure 12(a) does appear to indicate that the boundaries first increase, then slow to zero, and the rate of boundary formation decreases with time as the progress of solidification. The solutions in Figure 11(c) indicate that at the peak, the new area could be small to very large ~ $10^8$ $m^2/m^3$. For a large concentration of subboundaries, the subboundaries will become a significant part of the overall volume, and many even interact with each other in unanticipated ways [52, 96, 104,167]. The predicted order of boundary formation ($m^2/m^3$) using MEPR is similar to the boundary area per unit volume predicted by Jones for rapid solidification [167].

It should be noted here that although Figure 10 shows a sigmoidal shape, it is not the full S-curve (except for the error function fit). Mathematically, however, it is identical to an S-curve (CDF in Appendix A). It clearly shows that a rapidly solidified region initially displays a peak rate of grain or subboundary formation, which then diminishes over time, consistent with the amounts and rates shown in Figure 11. The atomization study shown in Figure 12(a) reveals features similar to the white zones observed in splat or ribbon cooling, called Zone A [115,167], which also form in melt-spun rapidly solidified ribbons and develop into a coarser zone, Zone B [115]. An illustration of boundary formation after nucleation (black regions on the periphery of the round particles) (one labelled A) and a rapid growth process that involves a change in morphology from boundary-free to subboundary containing regions. A change in facet angle can also change the boundary spacing [114], but facets are not expected to form at large interface velocities. Figure 12 (b) [142] shows the microstructure in ingot-type solidification, displaying globular equiaxed grains, when the overall process is slower, however still initially rapid, and leads to coarse dendritic or equiaxed conditions. The final bimodal structure in Figure 12 (b) could arise from macro-segregation [49], possibly with a secondary undercooling and secondary recalescence event (shown e.g., schematically in Figure 3(a) in Section 1) [47, 48, 48, 54, 59, 89, 142, 143, 167]. Here, entropy generation may follow the fully developed symmetric S-curve behavior for a transformation, with transitions in the grain-boundary rate, consistent with the MEPR principle.

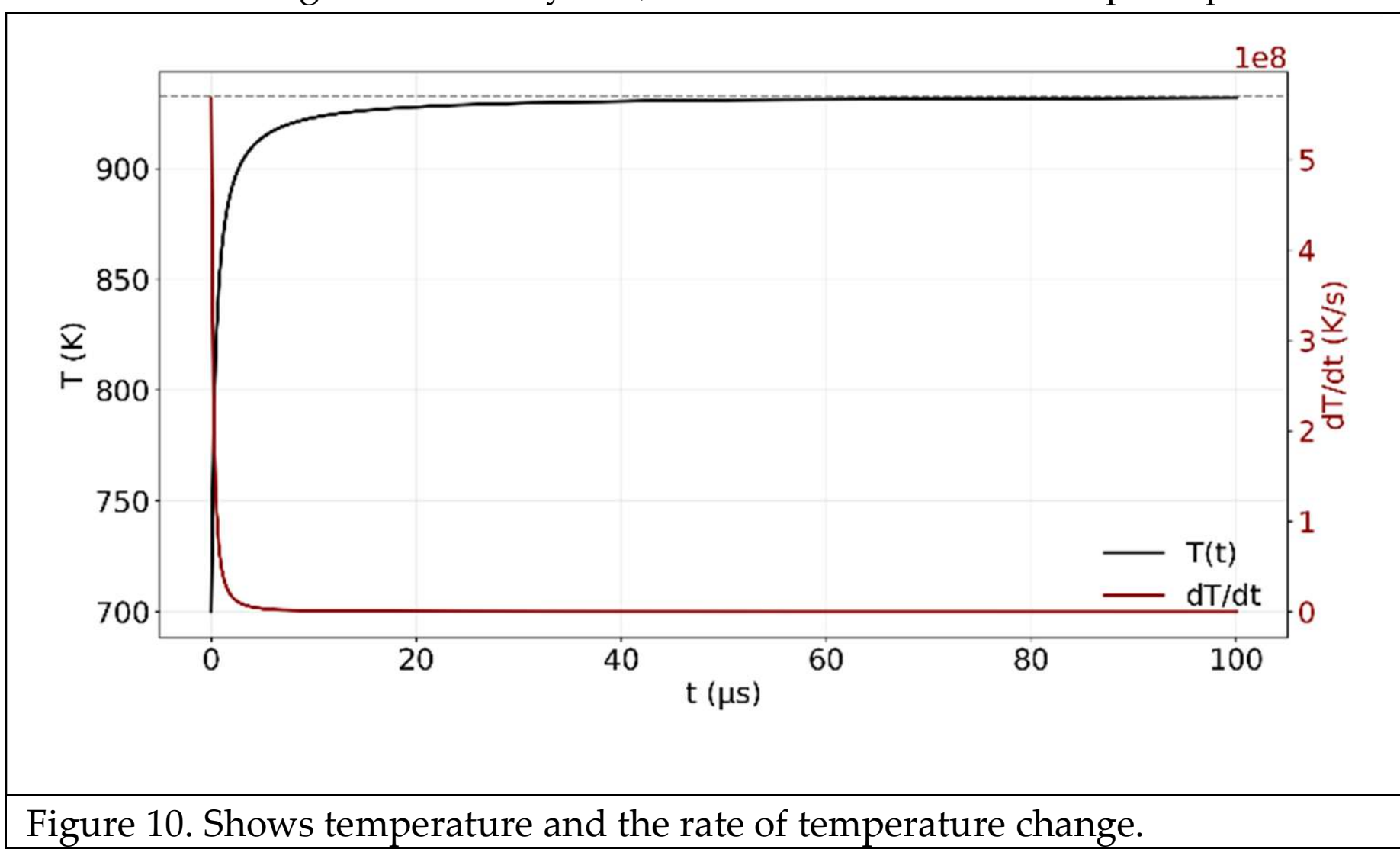

Figure 10. Shows temperature and the rate of temperature change.

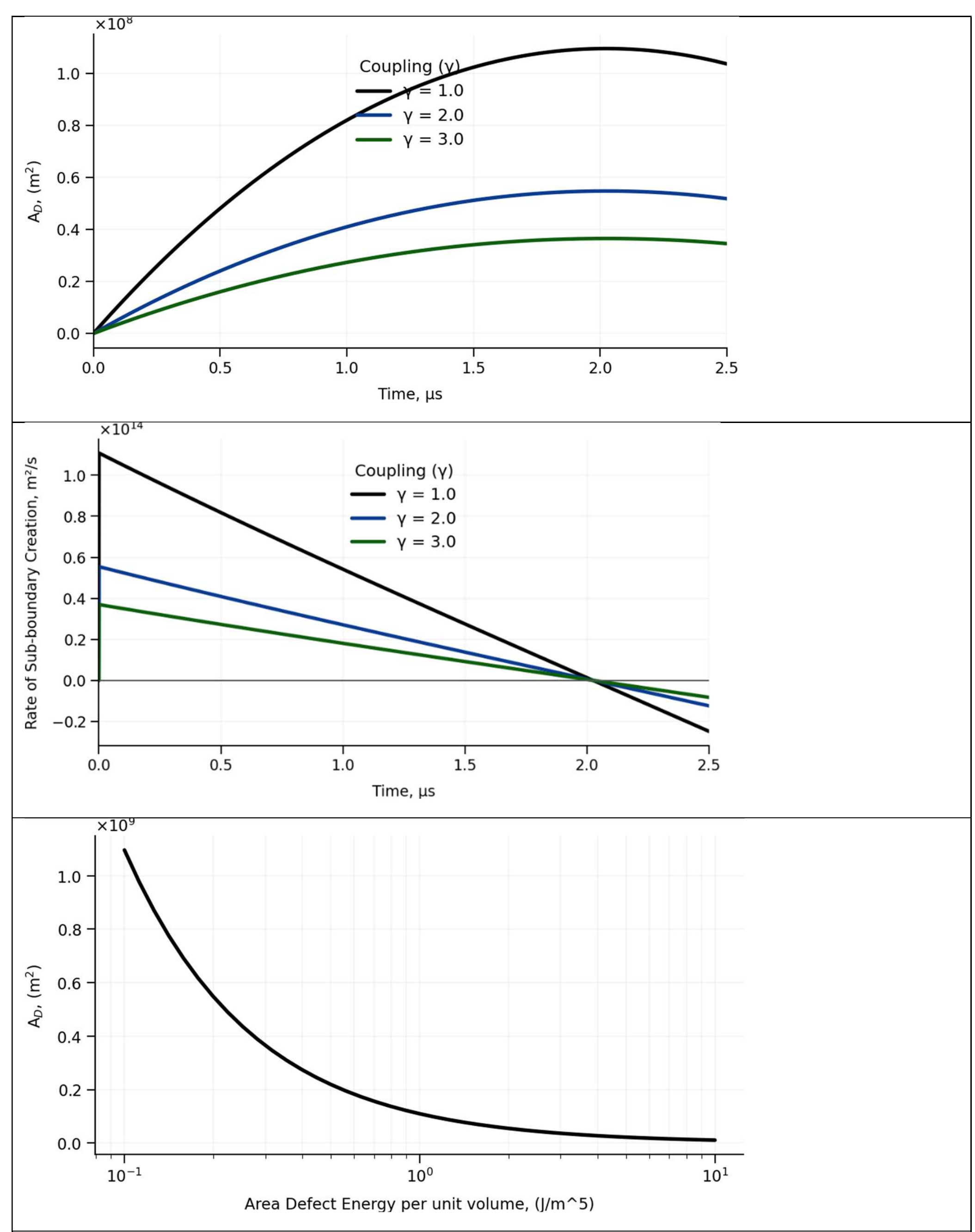

Figure 11. The MEPR condition yields the amount of sub-boundary area and the rate of sub-boundary formation as a function of time. $\Delta hsl = 1.6\times10^9$ J/m$^3$. γ varies as shown (J/m2). (a) is the subboundary area (per unit volume), (b) shows the rate of subboundary formation, and (c) indicates the maximum subboundary area formed as a function of γ (J/m$^2$). The graph in (c) is the variation of the maximum that is seen in (a) as a function of the boundary energy per unit volume.

## 5. The Significance of Sigmoidal-shaped curves (S-curves) for Sub-regions and Resilience.

In the processes studied above, we have been able to assert that (a) Processes at steady-state and non-steady-state tend to follow the principle of maximum entropy generation rate, (b) entropy generation can alongside work done (c) the transformation progresses according to a sigmoidal function for the fraction transformed and (d) for self-driven, self-organizing processes, the maximum entropy generation rate principle seems to govern the rates of transformation as it appears to match or at least follow physical observations

Sigmoidal (S-curve) transformation and related rates are ubiquitous in nature (see Figure 13). Many natural and artificial processes follow a sigmoidal (S-shaped) curve, characterized by slow initial growth, a rapid acceleration phase, and a final plateau as the process approaches its end. Autocatalytic reactions and forced reactions of the type discussed in this article often display S-shaped transformations. They are frequently linked to the formation of complex, self-organizing structures. Evidence presented in this article suggests that the process also maximizes the entropy generation rate, a key finding. Regardless of the processes, no process is isolated, and processes may have follow-on processes or alter the scale at which maximum entropy generation occurs [52], thereby triggering self-organization at a different scale for subsequent processes involving the same control volume. Figure 13 shows typical follow-on process paths.

When a metal is cold-worked, meaning it is deformed by a compressive process such as rolling to create sheet metal, the process increases the stored energy, thereby changing its relationship to its equilibrium state. Such a metal can be annealed (stress-relieved) by processes that redistribute energy among microstates. Both Recovery and Recrystallization are part of a self-organizing process called annealing. During recovery, dislocations reorganize within grains, relieving some internal stress. Recrystallization occurs quickly and discontinuously through a nucleation-like process, characterized by the formation and movement of high-angle grain boundaries. Finally, grain growth occurs in a manner like that described in Section 3 [131, 132, 135]. Each of these processes exhibits S-curve behavior individually, but may exhibit multi-S-curve behavior, which also resembles an S-curve, as shown in Figure 13. Figure 13 (b), redrawn from Reference [135], is an illustration of evolving properties with an S-Curve transformation during the process of recrystallization mentioned in the preceding paragraph. The figure illustrates the evolution of resilient properties such as Hardness and Ductility as a function of the fraction of recrystallized grains, which increases with time. Notably, a third process, known as Grain Growth [131, 132, 135, 136, begins to dominate as Recrystallization ends (this process has similarities to the sintering process discussed in Section 3. Note that, as pointed out in the caption of the figure, several S-curves that have different parametric constants for individual processes often interact (depending on the temperature of annealing) and show a single S-curve. On the left side of Figure 13(b), the starting state consists of metal grains that have undergone deformation and an earlier-stage entropy-generating process, Recovery and Recrystallization, during which boundaries formed, followed by a grain-growth process in which boundaries diminish (such processes can be modeled in the same manner as shown in Figures 11(b) and (c). Figure 13(c) illustrates metal creep, a phenomenon in which the control volume undergoes permanent shape changes under high temperatures and small anisotropic stress [131, 132]. Note the inverse S curves and inverse derivative curves as described in the figure caption. The strain rate vs. time exhibits an inverse sigmoidal rate behavior. Figure 13(d) is a schematic of a temperature-time profile that repeats over the course of several years for the diurnal cyclic human body temperature that varies gradually with age. Figure 13(d) is an illustration of a repetitive cycle of the S curve of the extender type, highlighted in Figure 13(a). Note that even for the cyclic example shown in Figure 13 (d), as discussed in Section 3, there are significant mathematical similarities between an S-shaped curve and a follow-on exponential decay.

The key point is that sintering, grain growth, or even the merging of soap bubbles are physical processes in which boundary areas, such as particle surfaces or grain boundaries, may increase or decrease during spontaneous self-organization. At first glance, this differs from the solidification and turbulence examples discussed earlier, in which new boundaries form during pathway selection, thereby helping to explain stored work or work done on a control volume during self-organization. However, the volume reduction during sintering absorbs the stored work involved in the process. Processes such as creep, recovery, and recrystallization can occur with or without applied stress. Heat can often be generated in such processes. The applied mechanical energy used to deform the material (e.g., forcing dislocations to move or climb over obstacles) is primarily converted into heat, minus a small fraction stored as microstructural defects. All such processes can be addressed using an entropy-generation equation, such as Equation 4.9. Any solution will thus require the MEPR condition to hold. Although this condition appears to be validated, confirmation has not yet been provided for every process.

Sigmoidal transformations are observed in chemical reactions for most coordinated growth processes across various dimensional scales. Examples include population growth patterns, like bacterial colonies expanding into environments with limited resources. In these cases, the growth rate initially accelerates and then slows down as it approaches the environment's capacity. Similarly, the reaction rate of allosteric enzymes or ligand binding, such as oxygen binding to hemoglobin, exhibits cooperative binding, in which the binding of one molecule facilitates subsequent binding. A notable example of S-curve behavior in earthquakes is the recovery of a community or infrastructure system after the event, as well as the planning and distribution of resources to enable faster, more effective responses [128]. Human diurnal temperatures also exhibit a sigmoidal phase throughout the cycle [129], shown in Figure 13. Even social and technological processes, like the diffusion of innovations, follow an S-curve pattern. The S-curve indicates slower activity in the initial and final stages, with peak activity during the main phase. Sigmoid functions are commonly used as activation functions in neural networks to introduce nonlinearity and map input values to the [0, 1] range. S-curves feature a phase called the Lag (see Figure 13). This lag time isn't always clearly predictable, but it can be shortened. For example, a nutrient-rich medium can reduce the lag phase of microbial growth, and investments can substantially facilitate the diffusion of innovation [121,136,149,153,154,159].

Although an S-curve is commonly used to represent a self-organizing process, we note that skewness can also yield an S-curve-like transformation. By definition, a normal distribution has skewness 0. A skewness of 2 corresponds to an exponential distribution. Relevant plots for the standard normal (mean $\mu$, std. dev. $\sigma$, and $z = (x - \mu)/\sigma$) and for the exponential distribution ($\lambda$=1) ($\mu = 1$ and $\sigma = 1$ and $z = \frac{x-1/\lambda}{1/\lambda}$ are shown below. The distribution with the highest entropy depends on the constraints applied to it. For a real-valued distribution with a given mean and variance, the normal distribution has the highest entropy, representing the most "uncertain" distribution under those constraints. However, before the study reported in this article, it was unclear whether the entropy generation rate is maximized. It appears that, at least in the examples considered, an S-curve pattern emerges as a pathway in an entropy-maximization scenario. Every point along the sigmoidal curve, therefore, represents the maximum entropy distribution. The start and endpoints indicate the maximum entropy generation rate when a system's steady-state condition shifts to another, as the entropy rate demands change [44]. All entropy-generating processes always leave the system with a different energy distribution (and greater order) per unit volume after the transformation than before. In Section 1.2, it was noted that resilience (toughness) has units of $J/m^3$. The Full Width at Half Maximum (FWHM) of a distribution is a measure of ~76% of the transformation. Note from Appendix A: the FWHM is $\approx 2.35.\sigma$is the standard deviation.for a Gaussian distribution. For an Exponential distribution, FWHM is $\sim\ln(2) \times \text{Mean} \approx 0.693/\lambda$, indicating different dependencies on their width-defining parameters. This has implications for the energy distribution across feature groupings (states), but is not explored further in this article, except in the context of resilience. Thus, based on the results in Section 4, for a constant-energy process, self-organization can be understood as the redistribution of internal energy within the system, thereby affecting the final resilience. For completeness, it should be noted that the Entropy generation (PDF/CDF^2) cumulative plot is analogous to the *reverse hazard-like structure function* used to study system degradation, e.g., disease progression to death [158], which intuitively correlates the ratio with time-dependent decay as internal entropy increases.

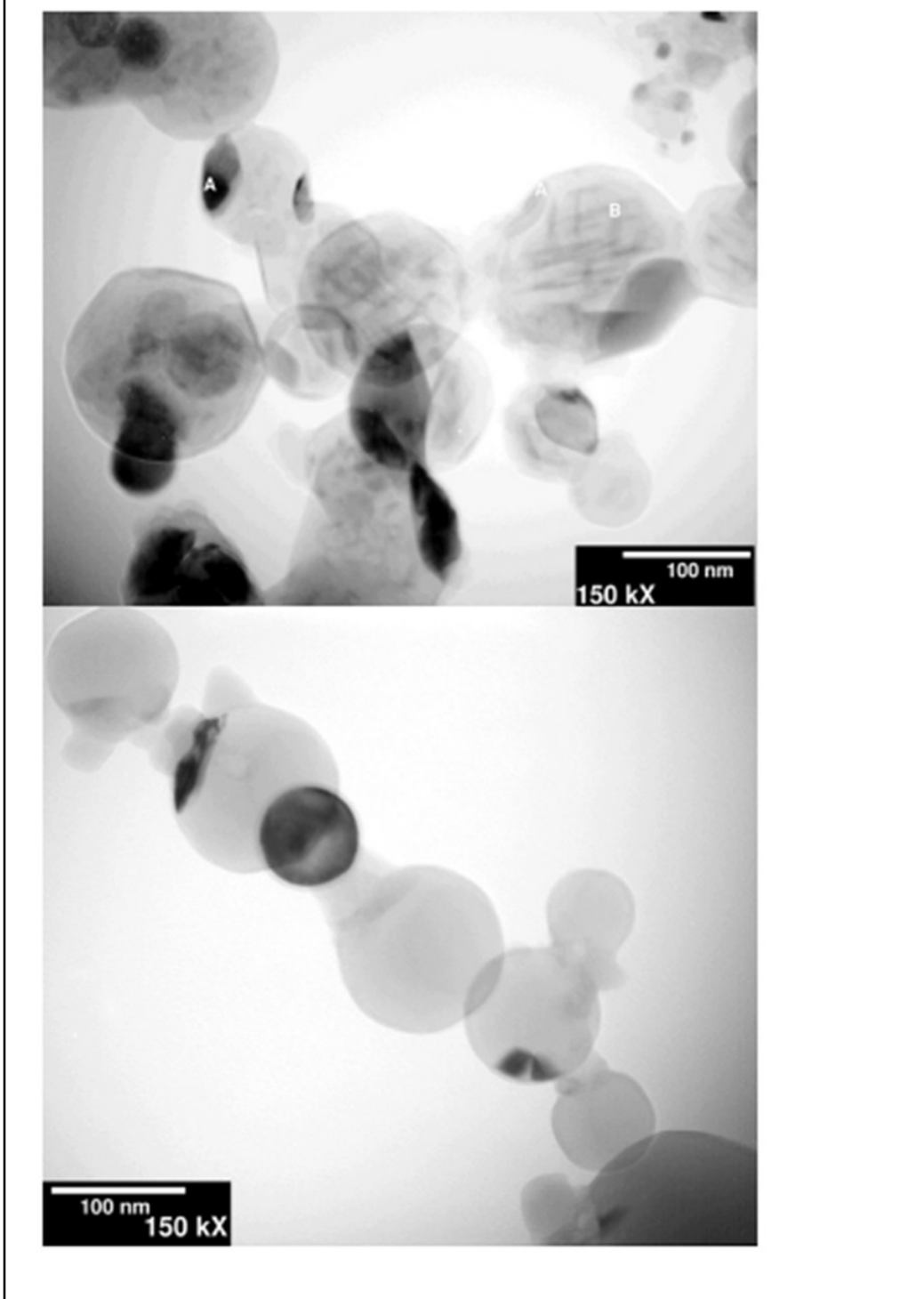

(a) Scale is nanometers

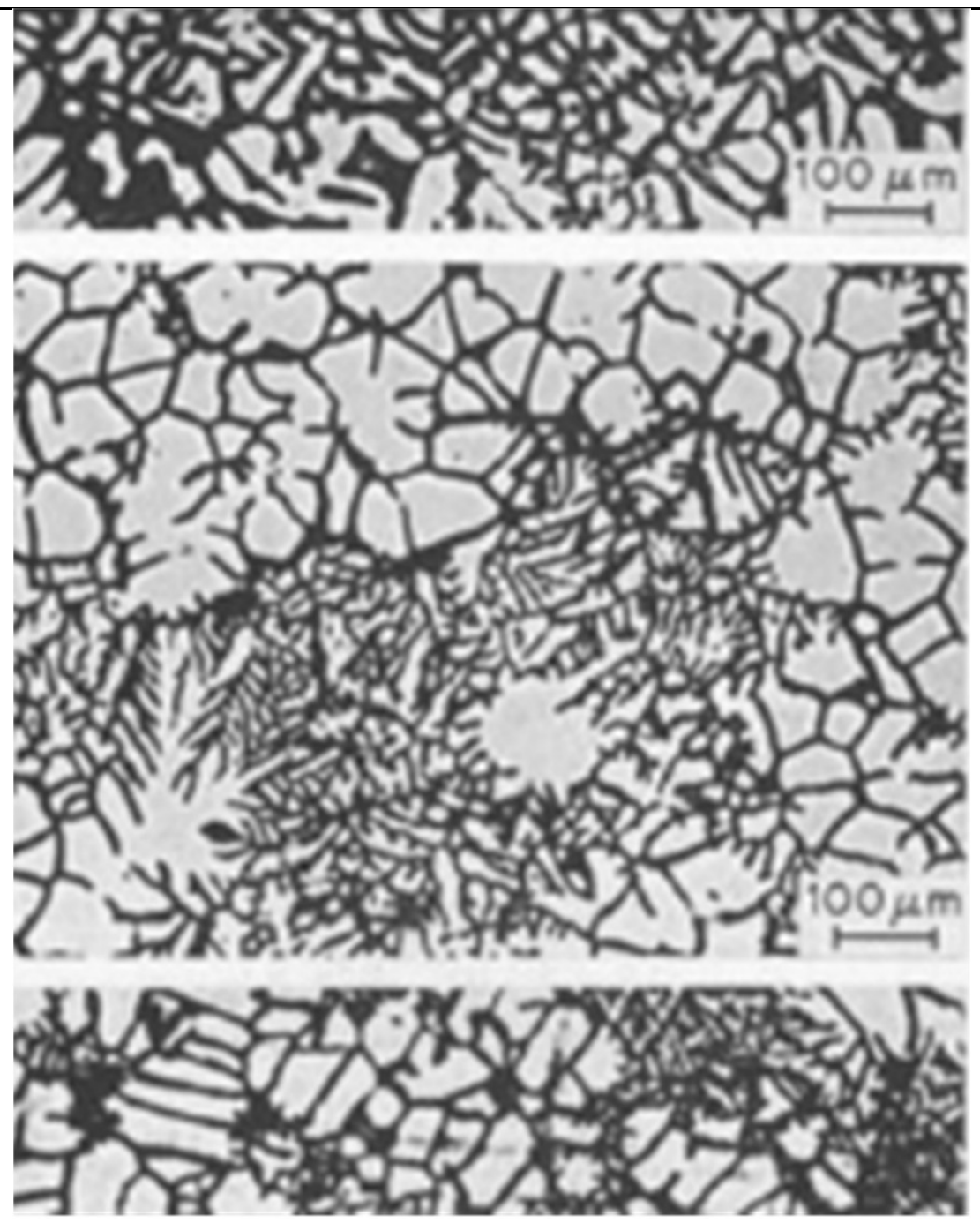

(b) Scale is micrometers to millimeters

Figure 12. (a) An illustration of boundary formation after nucleation (black regions on the periphery of the round particles) (one labelled A) and a rapid growth process that involves a change in morphology from boundary-free to subboundary containing regions. The Figure from reference [113] shows a TEM bright-field image of Al-Zn particles aged at 130 °C for 1 hour after an atomizing process. The aging process reveals the boundaries. Note the two types of transformations noted in the particle, namely precipitates (or a metastable phase) labeled **A** and the serrated structure labeled **B** (picture, white lettering): note that a dark region that encompasses the entire particle from a very tiny (50 nm), atomized particle. In the center of the photomicrograph is pure Zn, indicating that a highly nonequilibrium phase formed during the early stages of recalescence. Note the facets in this particle. (b) Shows bimodal sub-boundary envelopes following a path of metastable recalescence events or macrosegregation [47,48,49,143,167].

## 6. Features of Complex Self-Organization.

Self-organizing processes are complex due to pathway selection, the scale of entropy generation, property changes, and the timing (kinetics) of organization. They may require the simultaneous formation of multiple subregions and boundaries, with energy distributed through morphological adaptations that conserve energy within boundaries. For example, plants allocate energy to survival, resulting in reduced biomass, with stomatal transpiration regulated by light, humidity, and CO2 [126, 137]. Similar trade-offs occur in climate predictions, balancing wind and cloud types [44]. In metal creep, energy partitioning at grain boundaries helps assess embrittlement [136], often based on the quality and quantity of subregions. This affects the evolution of body armor in ants [146]. Different subregions coexist in large, slow-changing canopy systems, such as the human body or brain, though these are rarely modeled due to their complexity [52, 100-102, 145-146, 150]. This multi-region coexistence was noted early by Carnot and Clausius using control-volume concepts. The article does not examine how pattern subgrouping affects energy efficiency or resilience, but recent studies suggest that transport pathways can be optimized [145, 146, 150, 151].

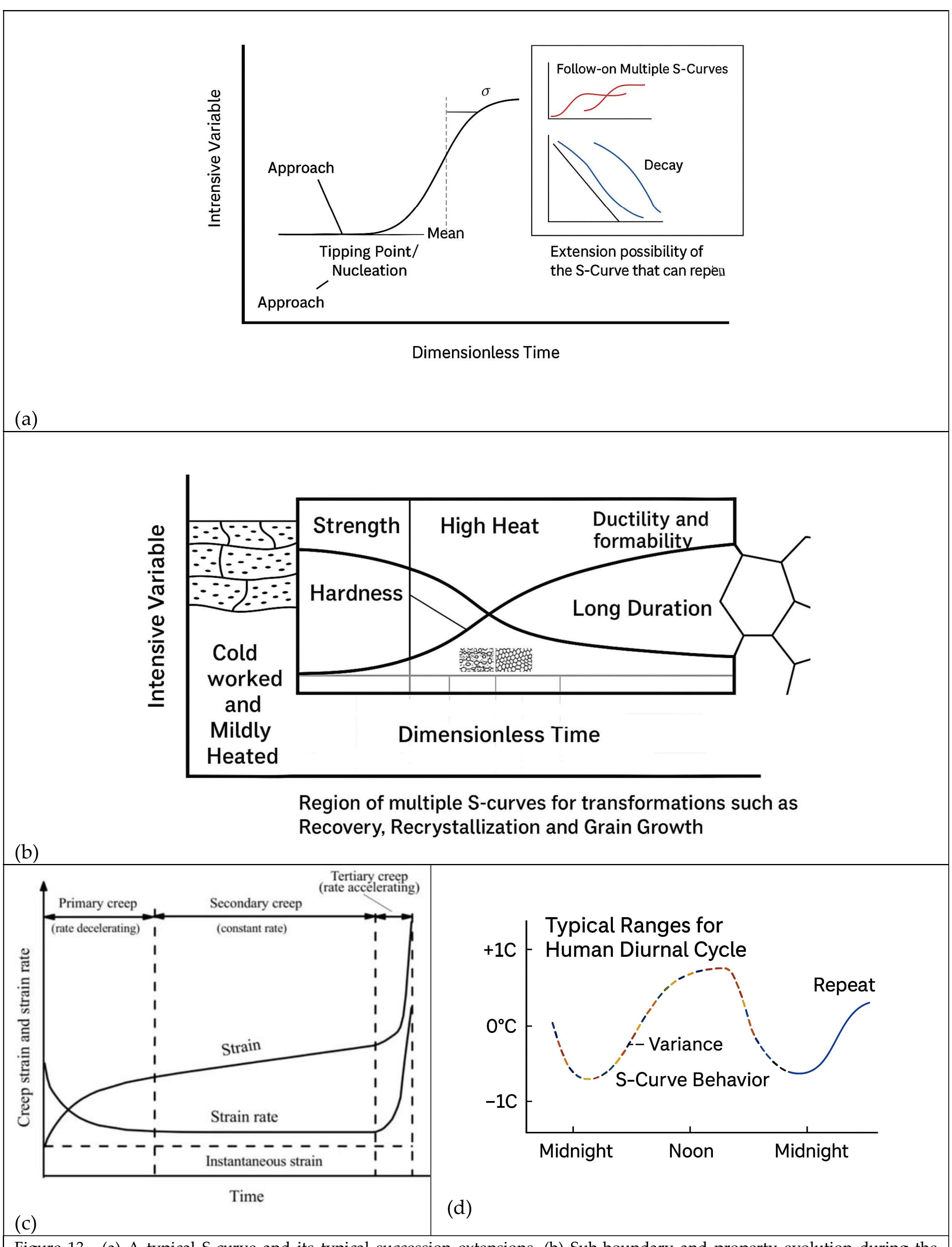

Figure 13. (a) A typical S-curve and its typical succession extensions, (b) Sub-boundary and property evolution during the recrystallization, recovery and grain growth of metals [135], here the typical parametric constants for the individual process interact to define the parametric constraints for the overall S-Curve, (c) Inverted S-curves used to describe typical creep and strain rate during creep [131,132] (d) a typical plot of the variation of the human body temperature every 24-hour cycle, modified from reference [129] to indicate the S-Curve tendencies followed by a simple exponential decay for the extension of the S-curve and repeats.

When entropy is produced during an entropy-generating process, energy is redistributed within the system's high-entropy sub-boundaries or organized sub-regions, such as strain energy or other potential energies. These systems and components can directly and efficiently transfer energy as Work, as shown in Section 3 of this article. In all cases of entropy generation [33, 61, 62], there is potential to recover stored energy when it is transferred out of a control volume in an organized manner [118, 119, 120, 121, 126, 128, 129, 130, 131, 132, 136]. This energy remains accessible after passing through the entropy-producing self-organization pathway. Such processes can be simple, such as an elastic spring returning to its original shape after a thermomechanical process, such as insufficient annealing, or more complex, involving mass transfer out of a control volume. A classic example of a complex pathway is cloud-to-ground lightning, which occurs due to an entropy-demand rate [44]. Rainfall can be collected in high-potential-energy situations and stored or dammed to generate hydroelectricity or innovative methods [140]. Pathway choices are made for various reasons in many fields [4,5,110,146,148,150] that involve entropy generation and work within a control volume. These choices and sub-regions enhance resilience to deformation, reduce wasteful energy expenditure, and serve as barriers that can even prevent epidemics [110], as discussed in Sections 5 and 6. In some cases, very rapid energy release can cause fracture or reorganize boundaries, thereby enhancing resilience [111, 132, 133].

Internal energy redistributes within the control volume due to entropy generation. Previously, we saw how this energy can enable future work, such as energy storage or sub-boundaries, like birds in formation. Figure 11 indicates that the boundary-creation rate decreases during dynamic self-organization. The boundary itself could become stationary once steady-state conditions are reached. When this happens, e.g., during bird flight formation, entropy generation occurs with stored work, and birds absorb and store energy in some form including thermal energy. Power expenditure is regulated, and efficiency gains are observed.

Morphological choices for self-organization involve subregions and their boundaries that interact across a control volume and, in turn, respond to a tipping point by engaging with their surroundings. In the introduction, it was asserted that engineering properties evolve toward more resilient states through self-organization. Increasing the boundary area per unit volume enhances fracture resistance and similar properties. It can also boost hardness, i.e., resistance to dislocations, depending on the Hall-Petch constants. Reducing and reorienting the sub-boundary height improves resistance against creep. The creep rate in metals is proportional to the volume ($m^{-3}$) multiplied by stress for unstrained creep, like the term $\gamma.dA_D/dt$ as in Equation 4.7. However, there is a limit to decreasing the region size or enlarging the sub-boundary area, as indicated by the Kolmogorov and Trivedi limits (section 3) for turbulent flow and crystal growth, respectively. At these limits, new morphological and even new phase transitions may occur [52] as the demand for the entropy generation rate increases. In creep, grain rotation occurs, thereby altering grain-boundary energy. A similar situation also happens with non-equilibrium clouds, as discussed in Section 3.2. The smallest self-replicating structures may include regions of high energy density and high dissipation rates; this is observed, for example, as turbulence progresses from transitional states. During solidification, solute atoms often segregate at grain boundaries or between dendrite arms. This spatial rearrangement of chemical species constitutes a defect, thereby increasing the overall local entropy. When different phases or compositions are close to such boundaries, local mixing and non-uniformity generate entropy. In highly alloyed systems, the mixing entropy of multiple components, including eutectics and solute gradients, becomes a crucial factor in assessing the entropy production rate. In pipe flow, viscous dissipation at the small-eddy scale near the pipe wall is particularly significant in the entropy-generating process.

For a normal distribution with a time-varying mean, the distribution's normalized variance may be a martingale-type property, for example, leading to isotropic, single-valued diffusivity (m2/s). If the distribution remains symmetric, then there must be a conserved quantity associated with this. This parameter has units of (1/K). As a result, the change in free energy (the driving force) at any point during a process—whether at equilibrium or non-equilibrium—leads to a condition of no change in free energy (available useful work) or free energy rate (power), respectively. Non-equilibrium systems (discussed in Section 3) often operate at the maximum entropy production rate, meaning they convert available energy into heat as quickly as possible. For a phase transformation, this would correspond to the fastest possible conversion rate. Noether's reasoning [108], which involves symmetries, can be used to identify conservation laws, such as energy conservation. Symmetries are important because they lead to conserved quantities through Noether's theorem. The normal distribution is always symmetrical. However, not all symmetrical distributions are normal. Hermann and Schmidt [109] have studied systems completely bounded by external walls (i.e., a control volume). They define the "sum rules" imposed by the three types of dynamical displacements, which are satisfied. This is comparable in many respects to the zero-force condition for birds in V-formations [5], which optimizes V-formations for long-range power use.
The sub-regions and the new order that develops within them are not necessarily identical. Such groupings cause the optimization of overall properties, which often outweighs focusing on individual parts, as different parts in a body or system become active at different times, such as the kidneys during the day with higher blood pressure and filtration, and less active at night. These differences can affect temperature patterns, with some organs slightly warmer than the core and extremities cooler due to heat loss. The body maintains an average core temperature of approximately 37°C, but regional variations exist. Brain temperatures can exceed 40°C in some regions, influenced by factors such as age, sex, and menstrual cycle phase. These temperature differences impact the organization of self-similar patterns in contact. Resilience, a context- and scale-dependent property, emerges from energy storage and thermal experience, with patterns such as V formations in birds indicating resilience to exhaustion.[5,6, 14, 20, 25,27,37,43, 44, 52, 58, 78, 83, 98, 99, 120, 121, 125,126, 127, 128,130, 131, 136, 147, 159]. Boundary formation, which influences thermal and structural stability, occurs under various physical conditions and supports cognitive reserve and adaptation. Plant resilience entails reconfiguring boundaries and prioritizing survival overgrowth. The stability of morphological structures relates to stored energy, with properties such as yield strength and hardness strongly influenced by grain boundaries and structural limits. For example, in metallic glass formation at high cooling rates, which increases strength but reduces ductility.

## 7. Summary and Concluding Remarks

In physical systems, local fluctuations can trigger events that lead to the formation of a new order, accompanied by the emergence of subregion boundaries (subboundaries). Such self-organization can occur spontaneously, within a control volume, when it reaches a tipping point. Based on the results presented in this article and previous results indicating that a steady-state pattern-forming process operates at a condition of maximum entropy generation rate, it seems likely that entropy generation rate (MEPR) is maximized during self-organization.

This is particularly expected when the fraction transformed (or an intensive variable such as temperature) exhibits S-curve behavior over the self-organization period. The formation of sub-boundaries enables increased resilience and connectivity while still maximizing the entropy generation rate and contextual energy efficiency. Individual processes that are Sigmoidal in nature can be succeeded by new processes that exponentially decay or followed by other sigmoidal processes, thereby permitting connections between discrete events in time. This article attempts a unified framework for understanding self-organization in terms of entropy production, subboundary adjustments, and the emergence of resilience.

## Appendix

Although an S-curve can be used to represent a self-organizing process (see Sections 5 and 6), we note that skewness can also yield an S-curve-like transformation (albeit with a tail), as shown in Figures A1, A2, and A3. By definition, a normal distribution has skewness 0. A skewness of 2 corresponds to an exponential distribution. Relevant plots for the standard normal (mean $\mu$, std. dev. $\sigma$, and $z = (x - \mu)/\sigma$) and for the exponential distribution (λ=1) ($\mu = 1$ and $\sigma = 1$ and $z = \frac{x-1/\lambda}{1/\lambda}$) are shown below.

$$PDF\ (z)(Normal)\ e^{-\frac{z^2}{2}},\ \ CDF(z)(Normal) = \frac{1}{2}\left[1 + \text{erf}\ \left(\frac{z}{\sqrt{2}}\right)\right] \quad \text{(A.1)}$$

For an Exponential (λ=1) (so $\mu = 1$and $\sigma = 1$). The domain is $x \geq 0$, which means $z \geq -1$. (Exp(λ)) $\text{PDF}(x)(exponential) = \lambda e^{-\lambda x}, x \geq 0$ $\text{CDF}(x)(exponential) = 1 - e^{-\lambda x}, x \geq 0$

$$\text{Note that } \frac{\lambda e^{-\lambda x}}{(1-e^{-\lambda x})^2} = \frac{\lambda e^{-(z+1)}}{(1-e^{-(z+1)})^2},\ z \geq -1 \quad \text{(A.2)}$$

The PDF and CDF (S-shaped curves) for each distribution are shown in Figure A1. The (PDF/CDF)$^2$, which approximates the entropy generation rate, is shown in Figure A2. The cumulative integral is shown in Figure A3.

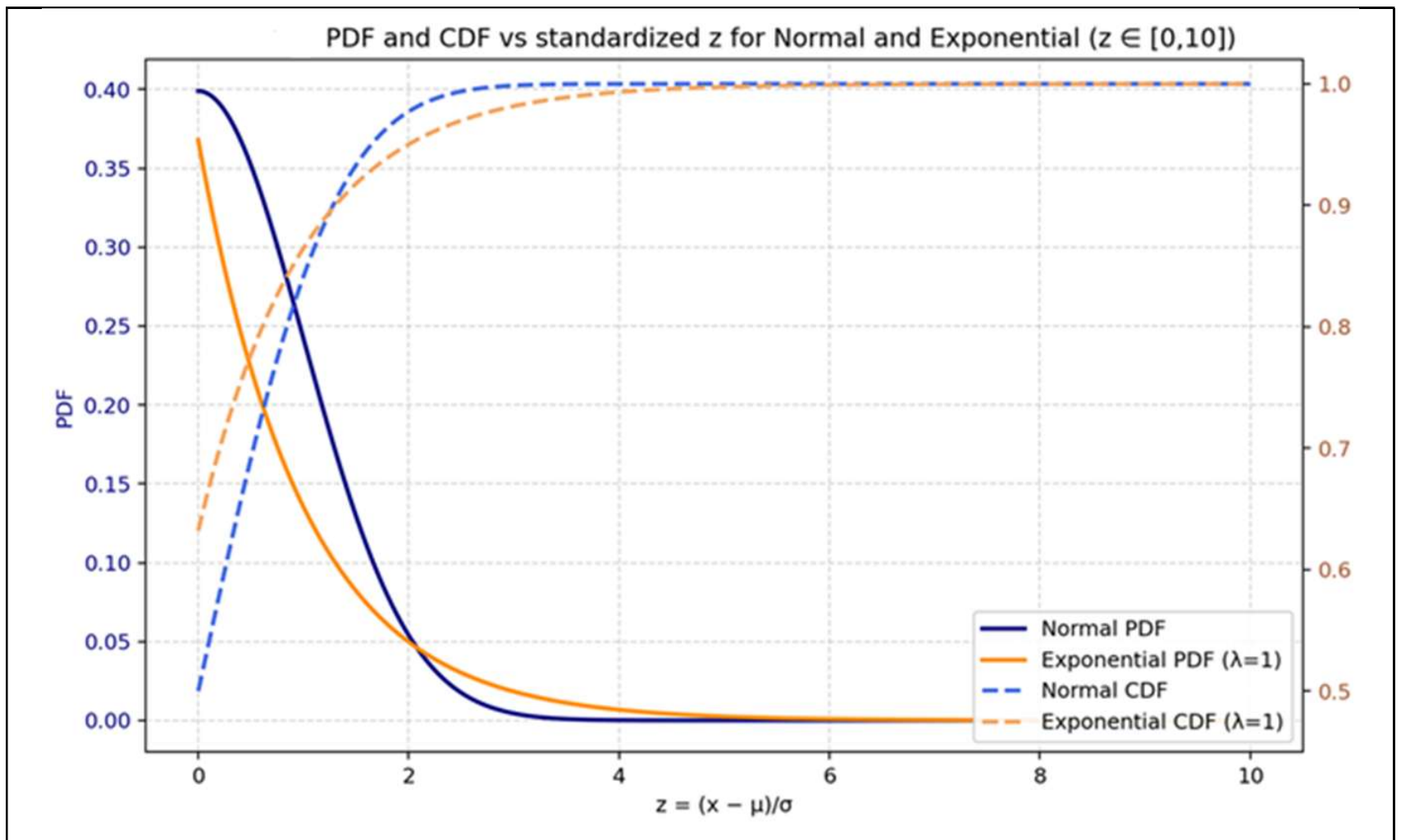

Figure A1. PDF and CDF vs. standardized Z, for Normal and Exponential functions (distributions)

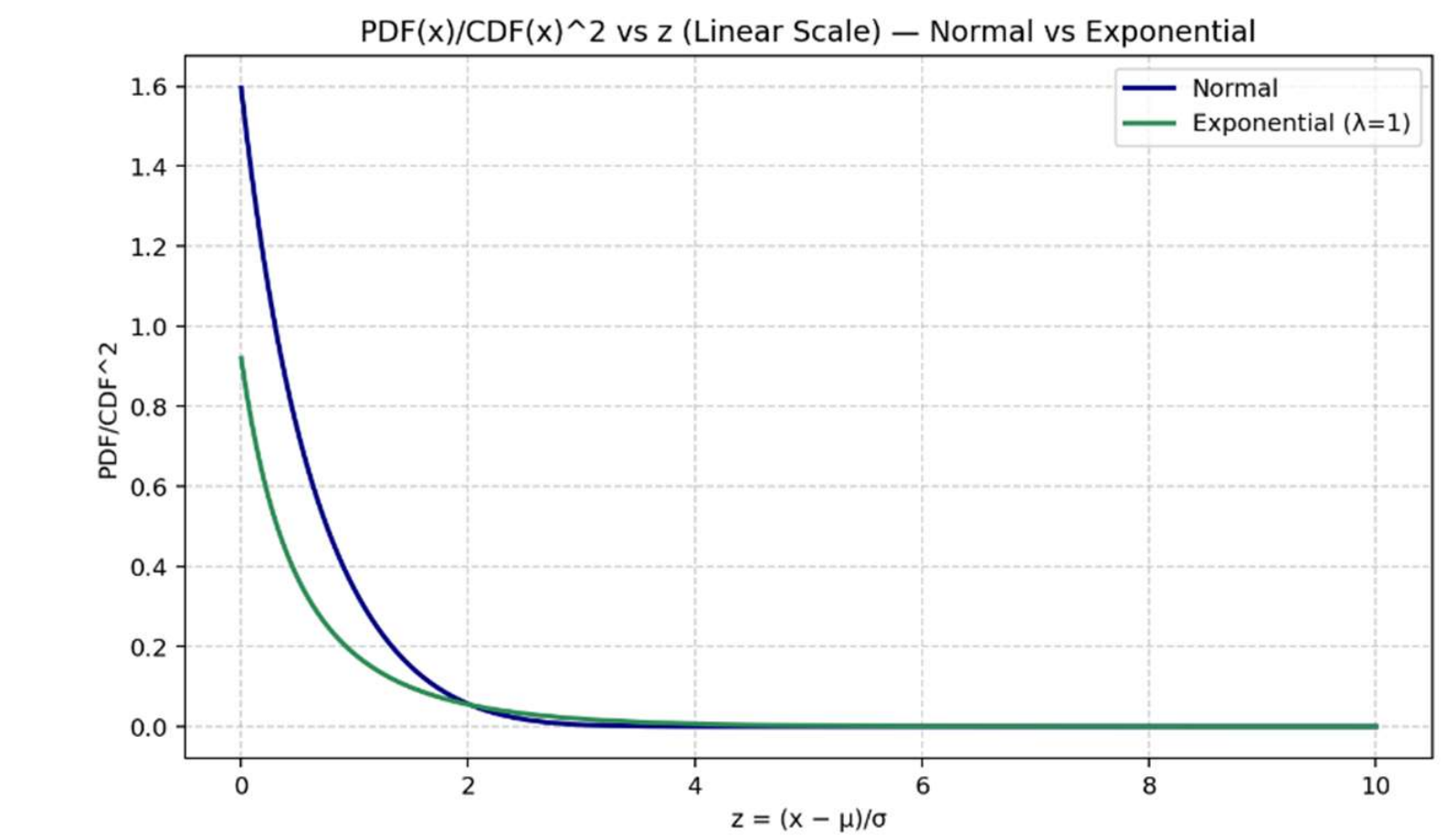

Figure A2: The approximate entropy generation rate for the Gaussian and Exponential functions

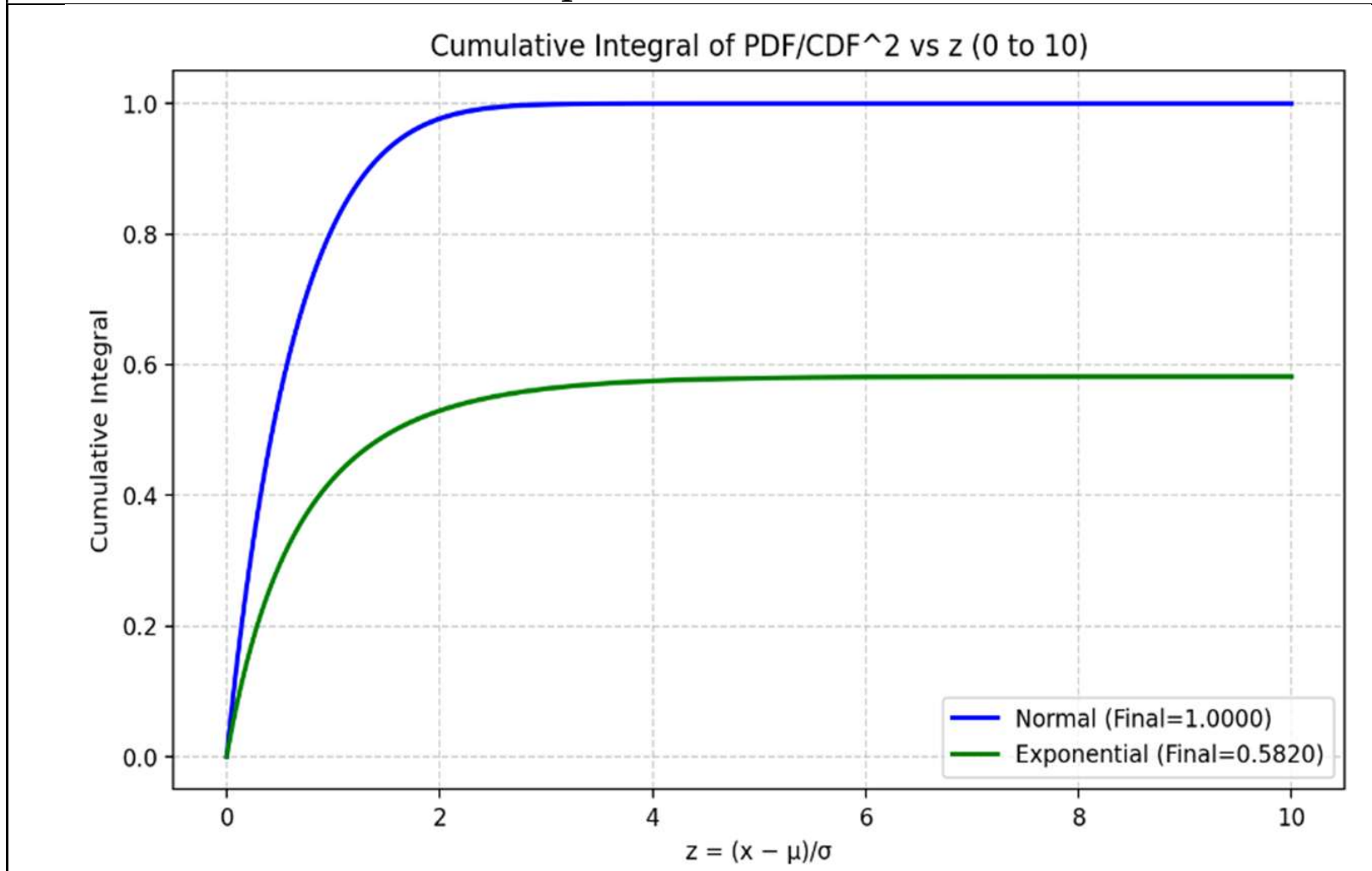

Figure A3: The approximated cumulative $(PDF/CDF)^2$, for the Gaussian and Exponential functions